\documentclass[onecolumn,aps,superscriptaddress,prd,10pt,showpacs,preprintnumbers,nofootinbib,eqsecnum]{revtex4-2}
\usepackage{graphicx,slashed,bbold,subfigure,amsmath,amssymb,amsthm,enumerate,color}
\usepackage[x11names,dvipsnames]{xcolor}
\usepackage{multirow}
\usepackage{soul}
\usepackage{enumitem}

\usepackage{natbib,ulem} 
\usepackage[colorlinks=true,citecolor=blue,linkcolor=blue,urlcolor=blue]{hyperref}
\usepackage{array}
\setcitestyle{square, numbers, comma, sort&compress}

\begin{document}

\title{Merging multidimensional equations of state of strongly interacting matter via a statistical mixture}

\author{Yumu Yang}
\affiliation{Illinois Center for Advanced Studies of the Universe\\ Department of Physics, 
University of Illinois Urbana-Champaign, Urbana, IL 61801, USA}

\author{Prachi Garella}
\affiliation{
 Department of Physics, University of Houston, Houston, TX 77204, USA
}

\author{Musa R. Khan}
\affiliation{
 Department of Physics, University of Houston, Houston, TX 77204, USA
}

\author{Tulio E. Restrepo}
\affiliation{
 Department of Physics, University of Houston, Houston, TX 77204, USA
}

\author{Joaquin Grefa}
\affiliation{Center for Nuclear Research, Department of Physics,
Kent State University, Kent, OH 44242, USA}
\affiliation{
 Department of Physics, University of Houston, Houston, TX 77204, USA
}

\author{Johannes Jahan}
\affiliation{
 Department of Physics, University of Houston, Houston, TX 77204, USA
}

\author{Mauricio Hippert}
\affiliation{Centro Brasileiro de Pesquisas Físicas, Rua Dr. Xavier Sigaud 150, 
Rio de Janeiro, RJ, 22290-180, Brazil}

\author{Jorge Noronha}
\affiliation{Illinois Center for Advanced Studies of the Universe\\ Department of Physics, 
University of Illinois Urbana-Champaign, Urbana, IL 61801, USA}

\author{Claudia Ratti}
\affiliation{
 Department of Physics, University of Houston, Houston, TX 77204, USA
}

\author{Romulo~Rougemont}
\affiliation{Instituto de F\'isica, Universidade Federal de Goi\'as, Avenida Esperan\c{c}a - Campus Samambaia, CEP 74690-900, Goi\^{a}nia, Goi\'{a}s, Brazil}

\begin{abstract}
We introduce a general method to merge multidimensional equations of state (EoSs) by combining them in a two-fluid equilibrium statistical mixture in the grand canonical ensemble. The merged grand potential density $\omega$ is built directly from the input EoSs and the fluid fractions are fixed by minimizing $\omega$ at fixed temperature $T$ and baryon chemical potential $\mu_B$. 
Thermodynamic consistency and stability are guaranteed as all thermodynamic quantities are consistently derived from a single merged grand potential $\omega(T,\mu_B)$ with the correct convexity properties. 
Our method can accommodate a first-order phase transition and a critical endpoint with mean-field critical exponents. 
We use this method to merge a van der Waals Hadron--Resonance--Gas  EoS with a holographic Einstein--Maxwell--Dilaton EoS that has a critical point and a first-order line. 
The result is a single EoS, spanning hadronic and deconfined matter over a broad range in $(T,\mu_B)$, which can be readily used in heavy-ion hydrodynamic simulations. Our merging method can be generalized to consider a higher dimensional phase diagram (e.g., by considering more chemical potentials) and more than two input EoSs.
\end{abstract}

\maketitle
\tableofcontents
\section{Introduction}
The equation of state (EoS) of quantum chromodynamics (QCD) plays a crucial role to determine the properties of matter under the extreme conditions found in astrophysical phenomena and nucleus-nucleus collision experiments  \cite{MUSES:2023hyz}. 
In particular, realistic EoSs covering a wide range of temperatures $T$ and baryon chemical potentials $\mu_B$ are important ingredients for the theoretical description of both binary neutron star mergers \cite{Baiotti:2016qnr} and relativistic heavy-ion collisions at beam energies probed by the RHIC Beam Energy Scan and, in the near future, by FAIR \cite{Luo:2015doi,Durante:2019hzd,Odyniec:2019kfh,Dexheimer:2020zzs,Senger:2021cfo,Almaalol:2022xwv,Sorensen:2023zkk}. 
However, no single theoretical framework is uniformly predictive across the entire QCD $T$ vs. $\mu_B$ phase diagram. 
Asymptotic freedom only renders perturbative QCD calculations reliable at energy scales that are typically much larger than those relevant for either astrophysical phenomena \cite{Fraga:2015xha,Annala:2017llu} or collision experiments \cite{Senger:2021cfo,Gyulassy:2004zy,Almaalol:2022xwv, Haque:2014rua,Alford:2007xm,Kajantie:2002wa,Vuorinen:2003fs,Laine:2016hma,Ghiglieri:2020dpq}.
Full non-perturbative QCD results are available via lattice QCD (LQCD) at $\mu_B=0$~\cite{Borsanyi:2013hza,Bazavov:2014pvz,Borsanyi:2025dyp,Borsanyi:2013hza}, but direct simulations at finite $\mu_B$ are hindered by the infamous sign problem \cite{deForcrand:2009zkb,Aarts:2015tyj}.
Many strategies have been developed over the years to circumvent this limitation, including Taylor expansion around $\mu_B/T=0$, reweighting techniques, and analytical continuation from imaginary $\mu_B$ \cite{deForcrand:2002ci,DElia:2002tig,Bonati:2015bha,DElia:2016jqh,Gunther:2016vcp,Bonati:2018nut,Borsanyi:2018grb,Bellwied:2015rza,Bollweg:2020yum,Borsanyi:2020fev,Bollweg:2022rps,Ding:2024sux}, as well as functional methods (see, e.g. \cite{Fischer:2018sdj,Fu:2019hdw,Gao:2020qsj,Bernhardt:2021iql}). Recently, new expansion schemes have extended first-principles control to somewhat higher $\mu_B$ \cite{Borsanyi:2021sxv,Borsanyi:2022qlh,Abuali:2025tbd}, for which the quantitative reliability typically holds up to $\mu_B/T\lesssim 3.5$. 
Nevertheless, results on a wider region of the QCD phase diagram will  in general rely on simplified and less systematic approaches, such as effective models --- each also restricted to some specific regime of applicability.  

In the confined regime, below the crossover temperature and at values of $\mu_B$ which are not too large, strongly interacting matter can be described in terms of its hadronic degrees of freedom. 
The hadron resonance gas (HRG) model \cite{Hagedorn:1965st,Vovchenko:2016rkn,Samanta:2017yhh,Karthein:2022zop} provides a simple but remarkably successful framework for this description. In its standard form, the HRG treats the system as an ideal gas composed of all experimentally known hadrons and resonances. To extend the model’s applicability toward finite baryon density, repulsive and mean-field effects can be incorporated via excluded-volume or van der Waals interactions, the latter reproducing the nuclear liquid–gas phase transition. Due to its quantitative agreement with LQCD thermodynamics up to and around the crossover region, the HRG model is widely employed to extract chemical freeze-out parameters from experimental hadron yield data, to initialize hydrodynamic and transport simulations, and to benchmark effective QCD models at low temperatures~\cite{Vovchenko:2019pjl,Kumar:2025rxj,Sasaki:2019pmp,PhysRevD.104.094508}. Nevertheless, the HRG model cannot capture the deconfined regime, where the relevant degrees of freedom are quarks and gluons, limiting its validity to the hadronic regime of the QCD phase diagram.

At higher temperatures, but below the perturbative regime, a strongly-coupled quark-gluon plasma (QGP) is found which cannot be accurately described in terms of hadronic degrees of freedom \cite{Bugaev:2008jj,PhysRevC.99.045204,Heinz:2013th,ALICE:2022wpn}. 
In this context, one may resort to holographic approaches capable of providing an effective framework to investigate both equilibrium and dynamical properties of the strongly interacting QGP~\cite{Kovtun:2004de,Casalderrey-Solana:2011dxg,Kovtun:2004de}, even though they cannot incorporate the hadronic degrees of freedom present in the HRG \cite{Rougemont:2023gfz,DeWolfe:2010he}.
A description, capable of reproducing LQCD results extrapolated to finite densities \cite{Borsanyi:2021sxv}, is provided by the bottom-up model developed in   
\cite{Critelli:2017oub,Grefa:2021qvt,Hippert:2023bel}, following the ideas from Ref.~\cite{DeWolfe:2010he}, in which  the QGP is described through black-hole solutions within a non-conformal Einstein-Maxwell-Dilaton (EMD) model in asymptotically Anti-de Sitter (AdS) spacetime, via the holographic gauge/gravity correspondence \cite{Maldacena:1997re} -- for a review of EMD models, see \cite{Rougemont:2023gfz}. The EMD potentials in this model were previously calibrated to continuum–extrapolated LQCD thermodynamics at $\mu_B{=}0$ and then used to predict the EoS at finite $\mu_B$, where the model naturally predicts a first-order transition line emerging from a critical point \cite{Critelli:2017oub,Grefa:2021qvt,Rougemont:2023gfz}. 
The parameter space for this model, as well as the location of the predicted critical point, was further constrained by a recent Bayesian analysis \cite{Hippert:2023bel} with good quantitative agreement obtained between the holographic equation of state and first-principles LQCD data at finite baryon density. 

Given the limited regime of applicability of individual approaches and the need for a unified description of strongly interacting matter across different phases, it is essential to construct an EoS that consistently covers a broad region of the QCD phase diagram. A comprehensive EoS should simultaneously describe the thermodynamics of confined hadronic matter and deconfined quark–gluon plasma, enabling realistic modeling of systems ranging from relativistic heavy-ion collisions to neutron stars and their mergers. Achieving this goal requires a controlled mechanism to merge or interpolate between distinct EoSs, each valid within its own regime, while preserving thermodynamic consistency, stability, and agreement with state-of-the-art LQCD results at both zero and finite baryon density. Such a framework should not only provide a physically consistent description from hadronic to deconfined matter but also allow for the inclusion of a critical point and a first-order transition line at finite temperature and density, offering valuable theoretical guidance for the experimental search for the QCD critical point and the interpretation of multi-messenger observations of neutron star mergers (when generalized to isospin asymmetric matter). Although several approaches have been proposed to merge different EoSs~\cite{Nonaka:2004pg,Parotto:2018pwx,Karthein:2021nxe,Kapusta:2021oco,Kapusta:2022pny,Kapusta:2025por}, most do not ensure thermodynamic stability, and agreement with state of the art LQCD thermodynamics is not always enforced.

In this work,  we develop a 
general method to interpolate between EoSs in a way that guarantees thermodynamic consistency and stability, and the possibility of a first-order transition line starting from a critical endpoint in the mean-field Ising universality class. 
This is achieved by modeling the system as an equilibrium statistical mixture of fluids which may interact with each other \cite{Callen1985_Thermodynamics,LandauLifshitz_StatPhys1,McQuarrie_StatMech,Huang_StatMech_2e}. 
In practice, we introduce a mixing weight $p$ that plays the role of an order parameter interpolating between the hadronic and QGP phases at $p=0$ and $p=1$, respectively. 
An EoS with the desired properties is then obtained by treating $p$ as an internal variable and setting it to its equilibrium value $p = \bar{p}(T,\mu_B)$.

As an application of this method, we merge a van der Waals hadron resonance gas (QvdW-HRG) EoS for the hadronic phase with the nonconformal holographic EMD EoS from Refs.~\cite{Critelli:2017oub,Grefa:2021qvt,Rougemont:2023gfz,Hippert:2023bel} for the QGP. 
The merged EoS %
enforces the correct limits and avoids spurious oscillations in derivatives by construction. Non-analytic features corresponding to the conjectured QCD first-order line and critical point are incorporated in a controlled way by an interaction energy between the two fluids ~\cite{Tokieda:2017vhm,Flory1942_JCP_ThermoHighPolymer,HildebrandScott1962_RegularSolutions,CahnHilliard1958_JCP_FreeEnergy,Goldenfeld1992_LecturesPhaseTransitions}. 
 Our construction is guided by continuum–extrapolated LQCD observables at $\mu_B{=}0$ and then extended to finite density by combining the holographic and QvdW-HRG descriptions, resulting in a single global description~\cite{Glendenning:1992vb,Masuda:2012ed,Alford:2013aca}. We obtain a single, consistent and differentiable EoS spanning up to $\mu_B\sim 1000\;\rm MeV$ and $T\sim 600\;\rm MeV$, in agreement with the state-of-the-art LQCD results at small values of $\mu_B$, where they are available. The resulting EoS and the open-source code for the merging procedure presented in this work are publicly available at \cite{yang_2026_18176810,pelicer_2025_17584997} and within the MUSES framework. Thus, other groups can readily use our new method to merge their favorite models for the different phases, and investigate the common features that emerge for the QCD phase diagram. 

The manuscript is organized as follows. In Section \ref{Sec:Merging} we present our new merging procedure, explaining how a crossover vs. first order transition can be implemented, and discussing how the thermodynamic quantities can be obtained. In Section \ref{Sec:Models} we introduce the two models for which we merge the EoSs, namely the QvdW-HRG EoS and the holographic EMD EoS. Results are presented in Section \ref{Sec:Results}, and we discuss our conclusions in Section \ref{Sec:Conc}.

\section{Merging Procedure
\label{Sec:Merging}}

{
Before outlining our new method, we clarify the issues with a commonly used strategy to interpolate between two equations of state \cite{Nonaka:2004pg,Parotto:2018pwx,Karthein:2021nxe,Kapusta:2021oco,Kapusta:2022pny,Kapusta:2025por}.

\subsection{Simple weighted average}

A simple way to continuously interpolate between two EoSs $P_1(T,\mu_B)$ and $P_2(T,\mu_B)$ in two different regimes of temperature and density is to take a weighted average of $P_1$ and $P_2$:
\begin{equation}
P(T,\mu_B)=S(T,\mu_B)\,P_1(T,\mu_B)+(1-S(T,\mu_B))\,P_2(T,\mu_B),
\end{equation}
where $S(T,\mu_B)\in[0,1]$ is a switching function. 
Thus, the two original EoSs are recovered as $S\to 1$ or $S\to0$ in their corresponding regions. This strategy has been applied in many different works \cite{Kapusta:2025por,Kapusta:2022pny,Nonaka:2004pg,Parotto:2018pwx,Karthein:2021nxe}, but it can lead to undesired artifacts and even unstable behavior, as we explain below.

In this method, one starts by defining the merged pressure, so that thermodynamic consistency can be guaranteed by obtaining the entropy and baryon densities as suitable derivatives of the merged $P(T,\mu_B)$.%
\footnote{Here, by ``thermodynamic consistency" we mean that the pressure $P(T,\mu_B)$ is the fundamental convex potential: its Legendre structure holds, i.e. $s=\partial_T P$, $n=\partial_{\mu_B} P$, $dP=s\,dT+n\,d\mu_B$, and its Hessian with respect to $(T,\mu_B)$ is positive semidefinite (Maxwell relations are satisfied; heat-capacity and susceptibility matrices are nonnegative).}
The issue then lies in the contributions of derivatives of the switching function to thermodynamical quantities. 
It is desired that the switching function $S(T,\mu_B)$ remains constant at $S=0$ or $S=1$ over the respective regimes of validity for each equation of state, and that it continuously switches between these two values in an intermediate window in $(T,\mu_B)$. Therefore, we expect $S$ to always cross an inflection point between these two regimes, where at least one of its derivatives has a maximum. This may lead to artifacts in derivatives of $P(T,\mu_B)$---for instance, in the baryon density:
\begin{align}
\begin{split}
    n(T,\mu_B)={}&\frac{\partial P}{\partial \mu_B}(T,\mu_B)\\
={}& S(T,\mu_B)\,n_{1}(T,\mu_B)+(1-S(T,\mu_B))\,n_{2}(T,\mu_B) + (P_1(T,\mu_B) - P_2(T,\mu_B))\frac{\partial S}{\partial \mu_B}(T,\mu_B),
\end{split}
\end{align}
where the first two terms are a weighted average of the densities in each EoS, $n_1$ and $n_2$, but the last term can lead to artificial contributions that peak at intermediate $T$ and $\mu_B$. 

More importantly, these contributions may lead to spurious contributions and oscillations in second derivatives of the pressure. For instance, for the second order baryon susceptibility, one finds 
\begin{align}
\begin{split}
    \chi_{2}(T,\mu_B)=\frac{\partial n}{\partial \mu_B}(T,\mu_B)
={}& S(T,\mu_B)\,\chi_{2,1}(T,\mu_B)+(1-S(T,\mu_B))\,\chi_{2,2}(T,\mu_B) 
\\
&+ (n_1(T,\mu_B) - n_2(T,\mu_B))\frac{\partial S}{\partial \mu_B}(T,\mu_B) 
\\
&+
(P_1(T,\mu_B) - P_2(T,\mu_B))\frac{\partial^2 S}{\partial \mu_B^2}(T,\mu_B),
\end{split}
\end{align}
where, again, the first two terms are a weighted average of the susceptibilities $\chi_{2,1}$ and $\chi_{2,2}$ for each EoS, and the terms on the second line are derivative terms, which are expected to peak in the intermediate region between the domain of validity for each description. 
On the third line in the equation above, the second derivative of $S$ can  lead to oscillations, which may turn the $\chi_2$ for the merged equation negative and generate a thermodynamic instability. 
Similar terms will be present in all second-order derivatives, jeopardizing the convexity of the pressure \cite{LandauLifshitz_StatPhys1} and, thus, the thermodynamic stability of the final merged EoS $P(T,\mu_B)$ defined using this approach.

In general, it is hard to guarantee without explicit numerical calculations that the contributions above will not cause unstable or unphysical behavior in a given application of this merging procedure. Below we provide an alternative method in which these issues can be handled.

\subsection{Switching via an internal variable}

The main novelty in our method is that we replace the prescribed switching function $S(T,\mu_B)$ by an internal variable $p\in [0,1]$, which may be interpreted as an order parameter that sets two different phases apart. 
Thus, in the grand canonical setting (fixed $T$ and $\mu_B$), equilibrium requires a local minimum of the grand potential density $\omega(T,\mu_B;p)=-P(T,\mu_B;p)$ in terms of $p$,
\begin{equation}
\label{eq:min_omega}
\frac{\partial\omega}{\partial p} (T,\mu_B;p)=0,\qquad \frac{\partial^2\omega}{\partial p^2} (T,\mu_B;p)\ge 0.
\end{equation}
The condition on the left eliminates contributions from $\partial P/\partial p$ to first derivatives of the pressure, while the one on the right avoids terms that could make the pressure non-convex. 

We then turn to the construction of the grand-potential density $\omega(T,\mu_B;p)$. 
If we define it as the weighted average $\omega = - P_2 - p(P_1-P_2)$, the free energy becomes an affine function of $p$, $\partial\omega/\partial p$ only vanishes for $P_1=P_2$
and, because an affine function has no local interior extrema (unless it is a constant), the minimum of $\omega$ will always be at the boundaries of the interval (either $p=0$ or $p=1$). 
Physically, this constraint means that a linear switching of the
pressure cannot interpolate between two genuinely distinct phases without
violating the Legendre structure of thermodynamics\footnote{In other words, a linear switch cannot produce an interior minimum between distinct phases at fixed $(T,\mu_B)$. Equilibrium selects the phase with larger pressure (with coexistence only if $P_1=P_2$).}. 
Therefore, we must add extra terms to $\omega$ which are nonlinear in $p$. 

In practice, starting from an explicit functional form for $p(T,\mu_B)$ and finding a corresponding form of $\omega$ that satisfies Eq.~\eqref{eq:min_omega} can be very challenging. Thus, we start from $\omega$ and minimize it to find $p$. 
For a functional form of $\omega$ that leads to the desired features, we take guidance from the physics of 2-fluid mixtures, with $p$ and $1-p$ representing the fractions of fluids with pressures $P_1$ and $P_2$, in the spirit of Flory-Huggins solution theory \cite{flory1953principles}. 
Our Ansatz for the grand-potential density is of the form,
\begin{equation}\label{eq:free_energy}
\omega = -p\,P_1 - (1-p)\,P_2 + a\,\left[p\,\ln p + (1-p)\,\ln(1-p)\right] + p\,(1-p)\,b.
\end{equation}
The term $a\left[p\,\ln p + (1-p)\,\ln(1-p)\right]$, with $a>0$, corresponds to a Shannon-type entropy contribution arising from the entropy increase in mixing the two EoSs \cite{jaynes1957information}, and favors a more balanced mixture. Such a mixing term ensures the convexity of the merged grand potential density, which is desirable but precludes the emergence of a critical point and a first-order phase transition. 
In order to create a critical point and a first-order phase transition, we introduce the term $p\,(1-p)\,b$, encoding repulsive interactions between the two fluids. 
The competition between these two terms creates a double well that will allow for a Maxwell construction in equilibrium. In principle, $a$ and $b$ are free functions that one can tune to control the contributions of the mixing terms, as long as $a > 0$. 

}

At any fixed $(T, \mu_B)$, the mixing weight $p$ is determined by minimizing the grand potential density of the system. 
Local minima can be found by solving%
\footnote{There could also be minima with $\partial \omega/\partial p \neq 0$ at the boundary of the interval for $p\in[0,1]$, but this is never the case for $a>0$, as can be seen upon close inspection of Eq.~\eqref{eq:free_energy}.}
\begin{subequations}\label{eq:stability}
\begin{equation}
    \left(\frac{\partial \omega}{\partial p}\right)_{T,\mu_B} = -P_1 + P_2 + a\,\ln\frac{p}{1-p} + (1-2\,p)\,b = 0, \label{eq:1st_order}
\end{equation}
   under the condition that
\begin{equation}  
    \left(\frac{\partial^2 \omega}{\partial p^2}\right)_{T,\mu_B} = \frac{a}{p\left(1-p\right)} - 2b >0.\, \label{eq:2nd_order}
\end{equation}   
\end{subequations}
We denote the equilibrium mixing weight, i.e.\ the global minimum of $\omega$, as $\overline{p}(T,\mu)$. 
After obtaining the equilibrium grand potential density, one can evaluate the thermodynamics of the system, namely the pressure $P$, entropy density $s$, and the net baryon density $n$,
\begin{subequations}
\begin{align}
    &P=-\omega(T,\mu_B;\overline{p})=\overline{p}\,P_1 + (1-\overline{p})\,P_2 - a\,\left[\overline{p}\,\ln \overline{p} + (1-\overline{p})\,\ln(1-\overline{p})\right] - \overline{p}\,(1-\overline{p})\,b, \label{pressure}\\
    & s =  \left(\frac{\partial P}{\partial T}\right)_{\mu_B} = \overline{p}\,s_1 + (1-\overline{p})\,s_2 - \left(\frac{\partial a}{\partial T}\right)_{\mu_B}\left[\overline{p}\,\ln \overline{p} + (1-\overline{p})\,\ln(1-\overline{p})\right] - \overline{p}\,(1-\overline{p})\left(\frac{\partial b}{\partial T}\right)_{\mu_B},\label{eq:entropy} \\
    & n = \left(\frac{\partial P}{\partial \mu_B}\right)_{T} = \overline{p}\,n_1 + (1-\overline{p})\,n_2 - \left(\frac{\partial a}{\partial \mu_B}\right)_{T}\left[\overline{p}\,\ln \overline{p} + (1-\overline{p})\,\ln(1-\overline{p})\right] - \overline{p}\,(1-\overline{p})\left(\frac{\partial b}{\partial \mu_B}\right)_{T}.\label{eq:density}
\end{align}
\end{subequations}
If we keep $a$ and $b$ constant, the entropy and density become a simple weighted average of the ones for each input EoS, with mixing weights $\overline{p}$ and $1-\overline{p}$. 
Nonetheless, the interpretation of the term proportional to $a$ as a mixing entropy suggests that it should contribute to the entropy density, and thus $a$ should be a function of temperature. 
The behavior of $a(T)$ with temperature will be crucial for the appearance of a critical point in the merged EoS, as we will see below.

It is important to note that, because we are maximizing the pressure, this procedure naturally gives a larger weight to the EoS with the highest pressure. 
This means that, when merging two pressures $P_1$ and $P_2$ over their respective ranges of validity $\mathcal{D}_1$ and $\mathcal{D}_2$, one must ensure that $P_1>P_2$ in $\mathcal{D}_1$, while $P_2>P_1$ in $\mathcal{D}_2$. Otherwise, the desired EoSs will not be reproduced in the appropriate limits. 
\subsection{Phase transition}

Let us now analyze the possibility of a phase transition in the merged EoS. 
First, note that, for $P_1=P_2$, $p =1/2$ becomes a solution of Eq.~\eqref{eq:stability}, regardless of the choices for $a$ and $b$. 
Intuitively, this makes sense, as both input EoSs are equally favored under this condition. 
Moreover, in this case, Eq.~\eqref{eq:free_energy} is symmetric under $p\to 1-p$, which means it has a $\mathbb{Z}_2$ reflection symmetry around the extremum  at $p=1/2$. 
When $(\partial^2\omega/\partial p^2)_{T,\mu_B} > 0$, this solution is a minimum for $P_1=P_2$ and this symmetry is smoothly realized. 
If $(\partial^2\omega/\partial p^2)_{T,\mu_B} < 0$, on the other hand, $p=1/2$ becomes a maximum separating two symmetric minima, interpreted as two competing phases. 
Hence, these two scenarios correspond to a smooth crossover and a first-order phase coexistence line. 
Finally, in between these two regimes, $(\partial^2\omega/\partial p^2)_{T,\mu_B} = 0$, which corresponds to a critical point. 
Considering the $\mathbb{Z}_2$ symmetry of the problem, and that the free-energy function is analytical, this critical point should be in the mean-field Ising universality class.\footnote{It can be explicitly checked that our method leads to the critical exponents of the mean-field 3D Ising model.} 

Ultimately, the presence of a phase transition in the merged EoS will be determined by whether the $\mathbb{Z}_2$ symmetry at $P_1=P_2$ is spontaneously broken or not, which in turn will depend on the values of the parameters $a$ and $b$, as follows:
\begin{enumerate}[label=\roman*)]
    \item If $b<2a$ or $b=0$, the solution of Eq.~(\ref{eq:1st_order}) is unique and $\left(\frac{\partial^2 \omega}{\partial p^2}\right)_{T,\mu_B}$ is strictly positive, provided that $a\geq 0$. In this case, the change of phases occurs through a crossover. 
    \item If $b=2a$, the minimum of the grand potential density is at $\overline{p}=1/2$, and $\left(\frac{\partial^2 \omega}{\partial p^2}\right)_{T,\mu_B} = 0$. The system is therefore located at the critical point.
    \item If $b>2a$, Eq.~(\ref{eq:1st_order}) admits three solutions, with $\overline{p}=1/2$ corresponding to the central solution, which is a maximum since $\left(\frac{\partial^2 \omega}{\partial p^2}\right)_{T,\mu_B} < 0$. Two minima are present, related via $p\to 1-p$ and with the same depth, signaling a first-order transition and the need for a Maxwell construction. 
\end{enumerate}
\begin{figure}[h!]
\centering
\includegraphics[width=.5\textwidth]{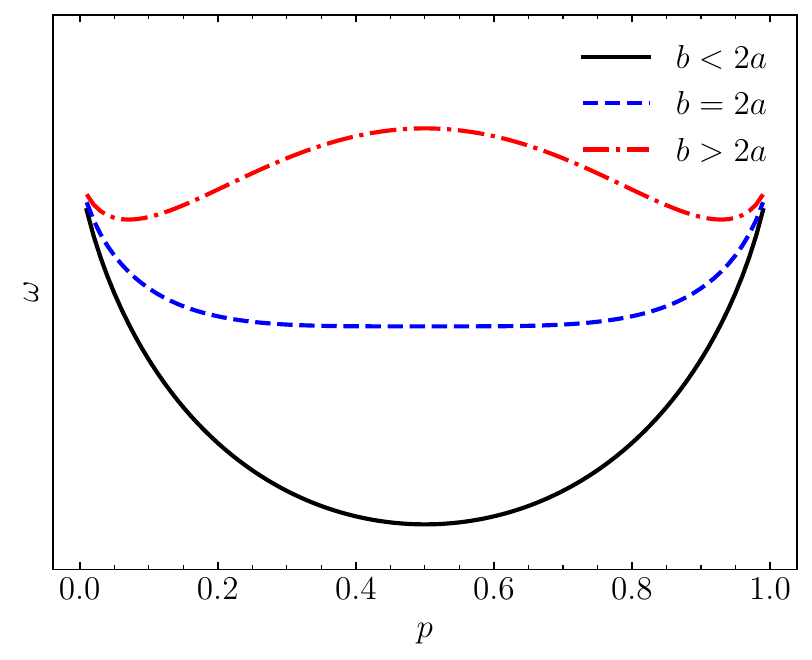}

\caption{Schematic figure of the thermodynamic potential when $P_1=P_2$, as a function of the probability $p$, for  three different scenarios: i) $b<2a$ (continuous black line), corresponding to a crossover; ii) $b=2a$ (dashed blue line), corresponding to the critical point; iii) $b>2a$ (dashed-dotted red line), corresponding to a first order phase transition.}\label{fig:potential}
\end{figure}
These scenarios are illustrated in Fig.~\ref{fig:potential}.

In the following, we show how our choices for $a$ and $b$ lead to a crossover at high temperature, which becomes a first-order line starting from a critical endpoint at a lower temperature.

\subsection{Thermodynamics}

Before we explore the thermodynamics of our merged EoS, we must choose the functions $a$ and $b$ that go into the free energy $\omega$. 
First, we note that the term proportional to $a$ represents an entropy of mixing, and should enter $s$ in Eq.~\eqref{eq:entropy}. We thus take  $a = T/\Delta V$, where $\Delta V$ is a parameter with dimensions of volume. 
The critical point for $b=2a$ is then found at the critical temperature $T_c = b\,\Delta V/2$, which leads to the natural definition of $b=2\,T_c/\Delta V$.
By plugging these definitions of $a$ and $b$ into Eqs.~\eqref{eq:stability}, one can check that there is always a stable global minimum of the free energy. If the pressures $P_1=P_2$ become equal at a temperature $T=T_t$, we find a crossover when $T_t>T_c$, and a first-order transition when $T_t<T_c$, as expected. 

With the choices above, we can calculate all the thermodynamic quantities from the grand potential density $\omega$. The net baryon density in Eq. (\ref{eq:density}) reduces to
\begin{align}
    n=\overline{p}\,n_1+(1-\overline{p})\,n_2,\label{eq:n}
\end{align}
 so that the thermodynamic relationship remains consistent with the statistical definition of number density \cite{Gorenstein:1995vm}.
The entropy density in Eq (\ref{eq:entropy}) reduces to
 \begin{align}
     s=\overline{p}\,s_1 + (1-\overline{p})\,s_2 - \frac{1}{\Delta V}\left[\overline{p}\,\ln \overline{p} + (1-\overline{p})\,\ln(1-\overline{p})\right] ,
 \end{align}
for which one naturally expects an increase in entropy from mixing different EoSs. 

Second derivatives of the pressure require derivatives of $\overline{p}$. By differentiating Eq. (\ref{eq:1st_order}) with respect to $T$ or $\mu_B$, we can obtain
\begin{subequations}\label{eq:p_deriv}
\begin{align}
    &\left(\frac{\partial \overline{p}}{\partial T}\right)_{\mu_B}=\frac{\overline{p}\left(1-\overline{p}\right)\Delta V}{T-4 T_c \overline{p}\left(1-\overline{p}\right)}\left[s_1-s_2-\frac{1}{\Delta V}\ln\left(\frac{\overline{p}}{1-\overline{p}}\right)\right], \label{dpdT} \\
    &\left(\frac{\partial \overline{p}}{\partial \mu_B}\right)_{T}=\frac{\overline{p}\left(1-\overline{p}\right)\Delta V}{T-4 T_c \overline{p}\left(1-\overline{p}\right)}\left[n_1-n_2\right].\label{dpdmu}
\end{align}
\end{subequations}
Using Eqs.~\eqref{eq:n} and \eqref{eq:p_deriv}, we can calculate the second order baryon number susceptibility, 
\begin{subequations}\label{eq:2nd_deriv}
\begin{align}
    \chi_2=\left(\frac{\partial n}{\partial \mu_B}\right)_{T}=\overline{p}\,\chi_{2,1}+\left(1-\overline{p}\right)\chi_{2,2}+\left(n_1-n_2\right)^2\frac{\overline{p}\left(1-\overline{p}\right)\Delta V}{T-4 T_c \overline{p}\left(1-\overline{p}\right)},\label{eq:chi2}
\end{align}
as well as the remaining second derivatives, 
\begin{align}
    &\left(\frac{\partial s}{\partial T}\right)_{\mu_B}=\overline{p}\left(\frac{\partial s_1}{\partial T}\right)_{\mu_B}+(1-\overline{p})\left(\frac{\partial s_2}{\partial T}\right)_{\mu_B}+\left[s_1-s_2-\frac{1}{\Delta V}\ln\left(\frac{\overline{p}}{1-\overline{p}}\right)\right]^2\frac{\overline{p}\left(1-\overline{p}\right)\Delta V}{T-4 T_c \overline{p}\left(1-\overline{p}\right)},\\
    &\left(\frac{\partial n}{\partial T}\right)_{\mu_B}=\overline{p}\left(\frac{\partial n_1}{\partial T}\right)_{\mu_B}+(1-\overline{p})\left(\frac{\partial n_2}{\partial T}\right)_{\mu_B}+\left(n_1-n_2\right)\left[s_1-s_2-\frac{1}{\Delta V}\ln\left(\frac{\overline{p}}{1-\overline{p}}\right)\right]\frac{\overline{p}\left(1-\overline{p}\right)\Delta V}{T-4 T_c \overline{p}\left(1-\overline{p}\right)}.\label{eq:dndT}
\end{align}
\end{subequations}
All of the derivatives above receive contributions from Eqs.~\eqref{eq:p_deriv}, which can be interpreted in terms of fluctuations from the statistical mixture of fluids, which are absent when $p=0,1$. As expected, these fluctuations diverge at the critical point, where $T=T_c$ and $\overline{p}=1/2$. 

 From the above, one may also compute the speed of sound and the specific heat at constant volume, respectively given by:
\begin{subequations}
\begin{align}
    &c_s^2=\frac{1}{\epsilon+P}\frac{ n^2 \left(\frac{\partial s}{\partial T}\right)_{\mu_B}-2\,s\,n\left(\frac{\partial n}{\partial T}\right)_{\mu_B}+s^2\chi_2}{\left(\frac{\partial s}{\partial T}\right)_{\mu_B}\chi_2-\left(\frac{\partial n}{\partial T}\right)_{\mu_B}^2},\\
    &C_V=T\left[\left(\frac{\partial s}{\partial T}\right)_{\mu_B}-\frac{1}{\chi_2}\left(\frac{\partial n}{\partial T}\right)_{\mu_B}^2\right].
\end{align}
\end{subequations}

\subsection{Ideal mixture}

In this paper, we focus on results for a mixture of interacting fluids with a critical point at $T_c>0$. However, it is instructive to examine the example of an ideal mixture, where the fluids do not interact and $b=0$. 
In that case, we can solve Eq.~\eqref{eq:1st_order} analytically to find
\begin{align}
\label{eq:idealp}
    \overline{p} = \frac{1}{1 + e^{-(P_1 - P_2)/a}}\;,
    & &
    1-\overline{p} = \frac{1}{1 + e^{(P_1 - P_2)/a}}.
\end{align}
We note that, for $a\to 0$, the above expression becomes a step function, $\overline{p} = \Theta(P_1 - P_2)$.  
The parameter $a$ therefore controls how smooth the crossover is between both input EoSs, which becomes a first-order transition for $a\to 0$. 

Substituting Eq.~\eqref{eq:idealp} into Eq.~\eqref{pressure}, we can also find the equilibrium pressure. Furthermore, replacing $a = T/\Delta V$, we find
\begin{equation}
    P = \frac{T}{\Delta V} \log \left( e^{P_1\Delta V/T} + e^{P_2\Delta V/T} \right).
\end{equation}
That is, for a volume $V=\Delta V$ we find a partition function $\mathcal{Z} = \mathcal{Z}_1 + \mathcal{Z}_2$, where $\mathcal{Z}= \exp(P\Delta V/T)$ and $\mathcal{Z}_i= \exp(P_i\Delta V/T)  $, with $i=1,2$. 
We thus interpret $(\Delta V)^{1/3}$ as the length scale for which a statistical mixture of fluids is allowed, which should be much larger than the microscopic scales of the problem. 
However, in this work, $\Delta V$ will be a free phenomenological parameter used in the construction of the merged EoS. 

Finally, by replacing $T_c=0$ in Eq.~\eqref{eq:2nd_deriv}, we find that extra fluctuations from the statistical mixture of EoSs scale as $\overline{p}(1-\overline{p})$, which is typical of the variance for a binomial distribution with success probability $\overline{p}$. 

\section{Models for the QCD equation of state\label{Sec:Models}}

\subsection{ Hadron Resonance Gas Model
\label{Sec:HRG}}
We model the confined low$-T$ sector with a hadron resonance gas that includes attractive and repulsive interactions via a van der Waals  equation of state. Following \cite{Vovchenko:2016rkn}, we include repulsive excluded-volume and attractive mean-field interactions between baryons but, unlike Ref.\ \cite{Vovchenko:2016rkn}, we also introduce excluded-volume repulsion between mesons (no attractive part). 
This is done to ensure the EoS is suppressed beyond its regime of applicability, i.e. the deconfined region of the phase diagram. Furthermore, baryon-antibaryon and meson-(anti)baryon QvdW cross terms are neglected in this work. With QvdW parameters fixed to nuclear-matter saturation properties (see below), this QvdW-HRG simultaneously (i) reproduces the nuclear liquid–gas transition at low $T$ and large $\mu_B$, and (ii) significantly improves the agreement with the susceptibilities at $\mu_B=0$ from LQCD, in comparison with the ideal hadron resonance gas \cite{Vovchenko:2016rkn,Vovchenko:2017zpj,Karthein:2021cmb}.
At $\mu_B=0$, mesons dominate bulk thermodynamics, so the inclusion of baryonic QvdW terms only modestly affects $P/T^4$ and $\epsilon/T^4$.\footnote{See Figs. 1--3 in Ref.~\cite{Vovchenko:2016rkn}.}

The total pressure contains contributions from repulsively interacting mesons $M$, QvdW-interacting baryons $B$, and antibaryons $\bar B$ as follows 
\begin{equation}
\begin{aligned}
P(T,\boldsymbol{\mu}) &= P_M(T,\boldsymbol{\mu}) + P_B(T,\boldsymbol{\mu}) + P_{\bar B}(T,\boldsymbol{\mu}),\\[4pt]
P_M(T,\boldsymbol{\mu}) &= \sum_{j\in M} P^{\mathrm{id}}_j\!\left(T,\mu^{*}_{j,M}\right),\\
P_B(T,\boldsymbol{\mu}) &= \sum_{j\in B} P^{\mathrm{id}}_j\!\left(T,\mu^{*}_{j,B}\right) - a\,n_B^{2},\\
P_{\bar B}(T,\boldsymbol{\mu}) &= \sum_{j\in \bar B} P^{\mathrm{id}}_j\!\left(T,\mu^{*}_{j,\bar B}\right) - a\,n_{\bar B}^{\,2},
\end{aligned}
\end{equation}
where $\boldsymbol{\mu}=(\mu_B,\mu_Q,\mu_S)$ and $\mu_Q$ and $\mu_S$ denote electric charge and strangeness chemical potentials.
The shifted chemical potentials (quantum QvdW in the grand canonical ensemble) are
\begin{equation}
\begin{aligned}
\mu^{*}_{j,M}      &= \mu_j - b_M\,P_{M},\\
\mu^{*}_{j,B}      &= \mu_j - b\,P_{B}      - a\,b\,n_{B}^{2}      + 2a\,n_{B},\\
\mu^{*}_{j,\bar B} &= \mu_j - b\,P_{\bar B} - a\,b\,n_{\bar B}^{2} + 2a\,n_{\bar B}\,,
\end{aligned}
\end{equation}
and the self-consistent densities are given by 
\begin{equation}
\begin{aligned}
n_{M}(T,\boldsymbol{\mu}) &= \bigl(1 - b_M\,n_{M}\bigr)\,
\sum_{j\in M} n^{\mathrm{id}}_{j}\!\left(T,\mu^{*}_{j,M}\right),\\
n_{B(\bar B)}(T,\boldsymbol{\mu}) &= \bigl(1 - b\,n_{B(\bar B)}\bigr)\,
\sum_{j\in B(\bar B)} n^{\mathrm{id}}_{j}\!\left(T,\mu^{*}_{j,B(\bar B)}\right).
\end{aligned}
\end{equation}
Since in the model we have $\mu_B\neq0$ and $\mu_Q=\mu_S=0$,
\begin{equation}
    n(T,\mu_B)=n_B-n_{\bar{B}}\,.
\end{equation}
Here, $P_j^{\mathrm{id}}$ and $n_j^{\mathrm{id}}$ are the ideal Fermi/Bose contributions for species $j$. The system of equations for $\{P_M,n_M\}$, $\{P_B, n_B\}$, and $\{P_{\bar B}, n_{\bar B}\}$ is solved iteratively at fixed $T$, $\boldsymbol{\mu}$, and thermodynamic derivatives (e.g. $s=\partial P/\partial T$, susceptibilities $\chi_n$) follow from $P(T,\boldsymbol{\mu})$. These equations are explicitly given in \cite{Vovchenko:2016rkn}.

Our implementation follows the setup of the QvdW-HRG of \cite{Vovchenko:2016rkn,Vovchenko:2017zpj}. 
For mesons, we include a modest purely repulsive term, with parameter $b_M>0$ 
and no attractive part, $a_M{=}0$. The numerical value is chosen  so that the total QvdW-HRG pressure does not increase too much at high temperature. 
For the (baryonic) parameters $a$ and $b$, we adopt the values
\begin{equation}
a = 329~\mathrm{MeV\,fm^{3}}, \qquad b = 3.42~\mathrm{fm^{3}},
\end{equation}
fixed by reproducing nuclear-matter saturation density $n_0=0.16\;\mathrm{fm}^{-3}$, and binding energy $E/A=-16~\mathrm{MeV}$ of the ground state of nuclear matter in the QvdW-HRG model. In this calibration, the liquid–gas critical point sits at $T_c \simeq 19.7~\mathrm{MeV}$, $ n_c \simeq 0.07~\mathrm{fm}^{-3}.$ As in the minimal QvdW-HRG~\cite{Vovchenko:2016rkn,Vovchenko:2015vxa,Vovchenko:2017zpj}, we apply the same $(a,b)$ to all baryonic species in the HRG.

We include all established strange and nonstrange hadrons from the Particle Data Group (PDG) listings (excluding the broad scalar states \(\sigma\) and \(\kappa\), as customary) %
and account for resonance widths by integrating over a relativistic Breit-Wigner mass distribution.

\subsection{Holographic Model \label{Sec:Holo}}

In the deconfined, but strongly-coupled regime, we model the thermodynamics of QCD with a five–dimensional gravity dual in which the quark-gluon plasma is represented by a charged black brane~\cite{Gubser:2008ny,DeWolfe:2010he,Rougemont:2015wca}.\footnote{Here, ``black brane'' denotes an asymptotically AdS$_5$ charged black-hole solution with a planar horizon (with $\mathbb{R}^3$ topology), appropriate as a gravity dual for a translationally invariant plasma in Minkowski $\mathbb{R}^{1,3}$. Following Ref.~\cite{Rougemont:2015wca}, the term ``black hole'' is sometimes used generically; however, the Ansatz solved there is explicitly a charged, spatially isotropic black brane. By contrast, a global AdS$_5$ black hole has a spherical horizon (with $S^3$ topology) and would correspond to the dual theory on a finite-volume three-sphere.} 
In particular, building on the principles of the gauge/gravity duality~\cite{Maldacena:1997re}, our holographic model implements a bottom-up, five-dimensional Einstein--Maxwell--Dilaton construction to describe hot and baryon-rich QCD thermodynamics~\cite{Rougemont:2015wca,Critelli:2017oub,Grefa:2021qvt,Rougemont:2023gfz,Hippert:2023bel}.\footnote{For other similar holographic EMD models which have been also recently applied to quantitatively study the physics of the baryon dense QGP, see e.g.~\cite{Knaute:2017opk,Cai:2022omk,Jokela:2024xgz,Chen:2024mmd,Zhu:2025gxo}.}
The bulk metric $g_{\mu\nu}$ sources the boundary stress-energy tensor where the extra holographic direction may be interpreted as a geometrization of the energy scale of the
renormalization group flow of the target gauge theory at the boundary.
The horizon encodes the temperature and entropy via Hawking’s relations, while a bulk $U(1)$ gauge field $A_\mu$ sources the conserved baryon number. The boundary value of its time component simulates the baryon chemical potential $\mu_B$, while the radial electric flux provides the baryon density $n$~\cite{DeWolfe:2010he,Hartnoll:2009sz}. 

Since QCD is intrinsically non-conformal due to its running coupling, the gravity dual must explicitly break conformal symmetry~\cite{Bazavov:2017dsy,Collins:1976yq,Gross:1973id,Politzer:1973fx}. In our holographic setup, this is achieved by introducing a dilaton field $\phi$ with a potential $V(\phi)$ that approaches AdS in the ultraviolet but deforms AdS in the infrared~\cite{Gursoy:2008bu,Gursoy:2007cb,Gubser:2008ny,Rougemont:2015wca}, thereby breaking scale invariance in the sector relevant for the QCD transition. The baryon sector is tuned via a gauge–dilaton coupling $f(\phi)$ so that the holographic $\chi_2$ matches the corresponding LQCD result at $\mu_B=0$~\cite{Grefa:2021qvt,Hippert:2023bel,Rougemont:2023gfz}.
The bulk gravity action is then given by
\begin{equation} \label{eq:gravity_action}
S=\frac{1}{16\pi G_5}\int d^5 x \sqrt{-g}\left[\mathcal{R}-\frac{1}{2}(\partial_\mu\phi)^2-V(\phi)-\frac{1}{4}f(\phi)F_{\mu\nu}F^{\mu\nu}\right],
\end{equation}
where $G_5$ is the 5-dimensional Newton's constant, $g$ is the determinant of the metric, $\mathcal{R}$ is the Ricci scalar, and $F_{\mu\nu}=\partial_\mu A_\nu -\partial_\nu A_\mu $ is the $U(1)$ field strength tensor.
Following \cite{Rougemont:2015wca}, we employ an asymptotically $AdS_5$ charged black-brane Ansatz (planar horizon) dual to an isotropic and translationally invariant plasma in $\mathbb{R}^{1,3}$,
\begin{align} \label{eq:ansatz}
ds^2 = g_{\mu\nu} dx^\mu dx^\nu= e^{2A(r)}\!\left[-\,h(r)\,dt^2 + d\mathbf{x}^2\right] + \frac{e^{2B(r)}}{h(r)}\,dr^2,\quad
\phi= \phi(r),\quad A_\mu = \Phi(r)\delta_{\mu}^{t} ,
\end{align}
whose horizon $r_h$ and boundary $r\rightarrow\infty$ map to finite $T$ and $\mu_B$ thermodynamics in the dual field theory.

From the action \eqref{eq:gravity_action} and the Ansatz \eqref{eq:ansatz}, one finds equations of motion which are numerically solved to obtain the thermodynamics of the model via the holographic dictionary. 
In practice, one computes the
entropy density $s$ and the baryon density $n$ over a grid in temperature $T$ and baryon chemical potential $\mu_B$ from the near-boundary, ultraviolet expansions of the EMD fields,  as
detailed in Refs.~\cite{DeWolfe:2010he,Critelli:2017oub,Grefa:2021qvt,Hippert:2023bel}. 
The pressure can be obtained by integrating the Gibbs-Duhem relation $dP = s\,dT+n\,d\mu_B$, starting from $\mu_B=0$ and some low reference temperature $T = T_0$, where the pressure is neglected.\footnote{We have checked that this leads to a thermodynamically consistent pressure which is independent of the integration contour.}\textsuperscript{,}\footnote{
 We obtain a finite result for the pressure by actually computing a pressure difference because our holographic model is still not renormalized. 
 Accounting for the pressure of the plasma by taking into consideration such a pressure difference is a reasonable approximation as long as $T$ is not close to $T_0$ (which lies deep into the hadronic phase, whose thermodynamics is not expected to be well described by holographic models~\cite{Rougemont:2023gfz} --- this is indeed one of the main phenomenological motivations for merging the EMD EoS with the QvdW-HRG EoS). 
 For a renormalized holographic EMD model, see~\cite{Cai:2022omk} --- one notes that the results for the pressure in the deconfined QGP phase of both EMD models are very similar, as both quantitatively agree with the first principles LQCD results.} The remaining thermodynamic functions (e.g. trace anomaly $I$, susceptibility $\chi_2$, speed of sound squared $c_s^2$, etc) follow from $P$, $s$, $n$ and their derivatives with respect to $T$ and $\mu_B$. 

For the dilaton potential and dilaton-gauge coupling, we take the Polynomial-Hyperbolic Ansatz of Ref.~\cite{Hippert:2023bel}:  
\begin{equation}
\begin{aligned}
V(\phi) = -12 \cosh(\gamma \phi) + b_{2}\phi^{2} + b_{4}\phi^{4} + b_{6}\phi^{6}\;,
\end{aligned}
\end{equation}
\begin{equation}
f(\phi) = \frac{\operatorname{sech}(c_{1}\phi + c_{2}\phi^{2} + c_{3}\phi^{3})}{1 + d_{1}}
+ \frac{d_{1}}{1 + d_{1}} \, \operatorname{sech}(d_{2}\phi).
\end{equation}
All model parameters are taken from the maximum \textit{a posteriori} values found in the Bayesian analysis from \cite{Hippert:2023bel,Hippert2024_ZenodoDataset}, where the model was constrained by continuum extrapolated LQCD results for $s$ and $\chi_2$ at $\mu_B=0$.
Results from this realization of the model are in quantitative agreement with the state-of-the-art LQCD results up to the largest available ratio of baryon chemical potential over temperature $\mu_{B}/T=3.5$, while predicting the emergence of a first-order phase transition line terminating at a critical point \cite{Grefa:2021qvt,Hippert:2023bel,Rougemont:2023gfz}. 

Using the EMD model to describe QCD thermodynamics makes the most sense in the region where the system is deconfined and most strongly interacting, as signaled by the trace anomaly, and the model calculations agree with state of the art LQCD at $\mu_B{=}0$, for $T>110\,\mathrm{MeV}$~\cite{Rougemont:2017tlu,Rougemont:2015wca,Rougemont:2023gfz,Bazavov:2017dsy,Borsanyi:2013bia}, and for moderate to high temperatures at finite $\mu_B$~\cite{Grefa:2021qvt,Hippert:2023bel}. The holographic model used here does not describe the confined hadronic phase (bound states, nuclear matter) at low $T$, nor is it intended to describe the high-$T$ region where asymptotic freedom becomes dominant ~\cite{Gubser:2008ny,Gursoy:2008bu,Gursoy:2007cb,Laine:2006cp,Bandyopadhyay:2015gva}. 
Thus, we consider the regime of validity of the holographic model to lie between those of the QvdW–HRG model and weak-coupling QCD \cite{Andersen:2011sf,Mogliacci:2013mca,Haque:2014rua}.

\section{Results \label{Sec:Results}}

Having defined our new general merging procedure, as well as the two input EoSs, we now apply our method to obtain a new EoS that  reproduces the QvdW-HRG and EMD models in the hadronic and QGP regimes, respectively. 
We choose the critical temperature to be $T_c = 126.1$ MeV, which leads to a critical baryon chemical potential of $\mu_c = 598$ MeV, consistent with the one predicted by the EMD model \cite{Hippert:2023bel}. 
The merger parameter $\Delta V$ must be adjusted for each specific pair of EoSs, as it depends on their individual pressure values and pressure difference between them. 
In the present case, we use $\Delta V=1.98 \times 10^{-6} $ $\rm{MeV}^{-3}$, which provides a smooth interpolation between the two descriptions in the crossover region. 

\begin{figure}[h!]
\begin{subfigure}
\centering
     \includegraphics[width=.45\textwidth]{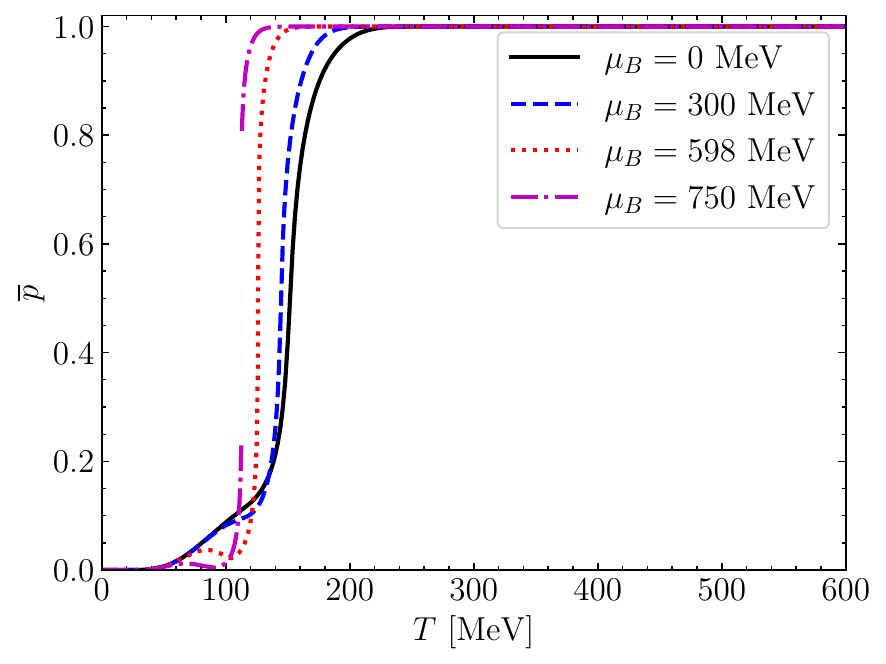}
\end{subfigure}
\begin{subfigure}
\centering
    \includegraphics[width=.45\textwidth]{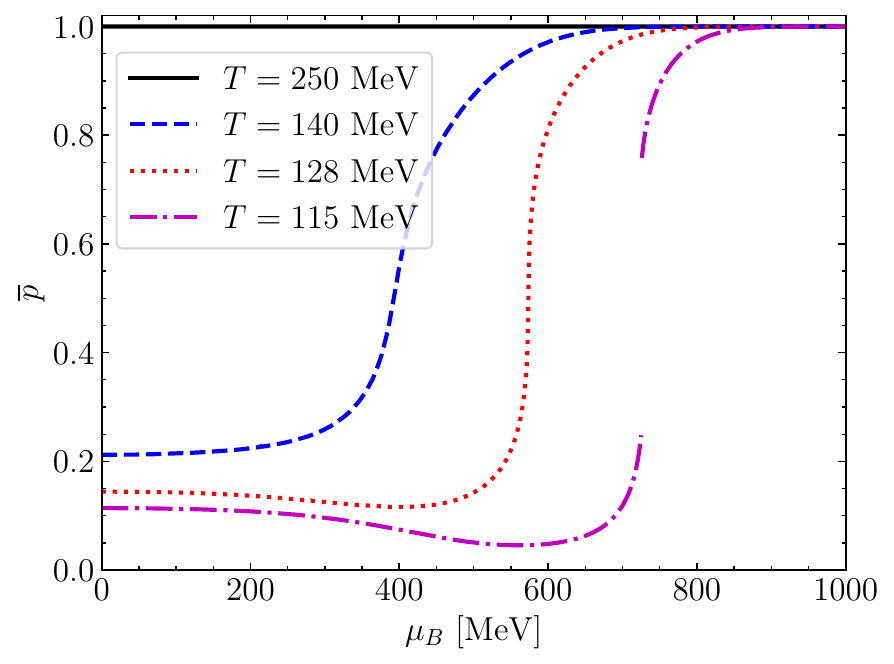}
\end{subfigure}
\caption{Mixing weight $\overline{p}$ of the EMD EoS as a function of temperature at different values of chemical potential (left) and as a function of chemical potential at different values of the temperature (right).}\label{fig:probability}
\end{figure}

Results for the mixing weight for the EMD model, $\overline{p}$, as a function of $T$ and $\mu_B$, are shown in Fig.~\ref{fig:probability}. 
As expected, $\bar{p}$ exhibits a discontinuity at $\mu_B>598$ MeV and $T<126.1$ MeV, which is a feature of a first-order phase transition, corresponding to an abrupt change between the hadronic and QGP descriptions. 
In contrast, at low $\mu_B$, $\bar{p}$ continuously transitions from 0 to 1 without any discontinuities, signaling a crossover.

\begin{figure}[h!]
\begin{subfigure}
     \centering
    \includegraphics[width=0.45\linewidth]{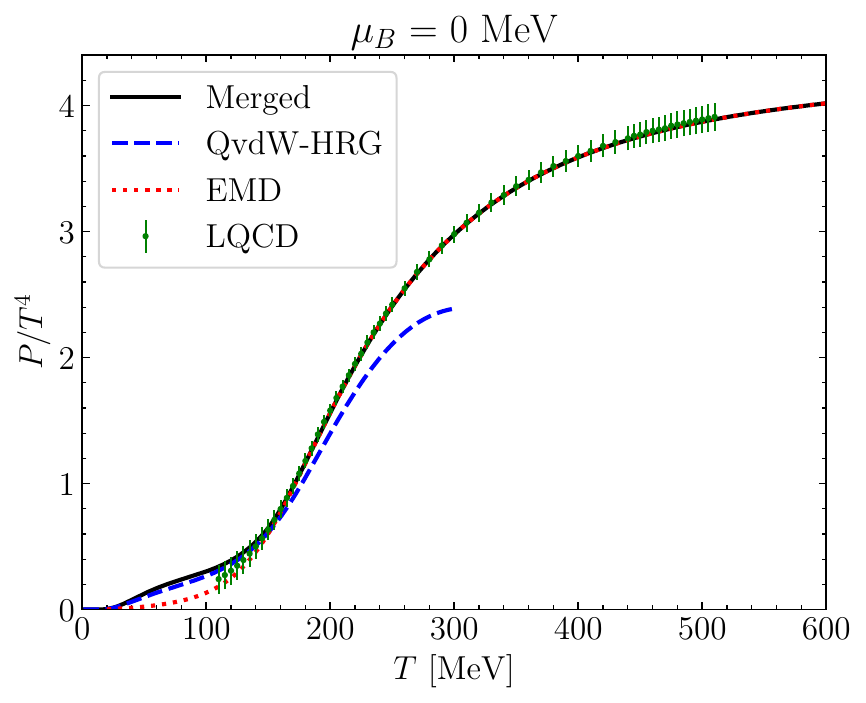}
\end{subfigure}
\begin{subfigure}
     \centering
    \includegraphics[width=0.45\linewidth]{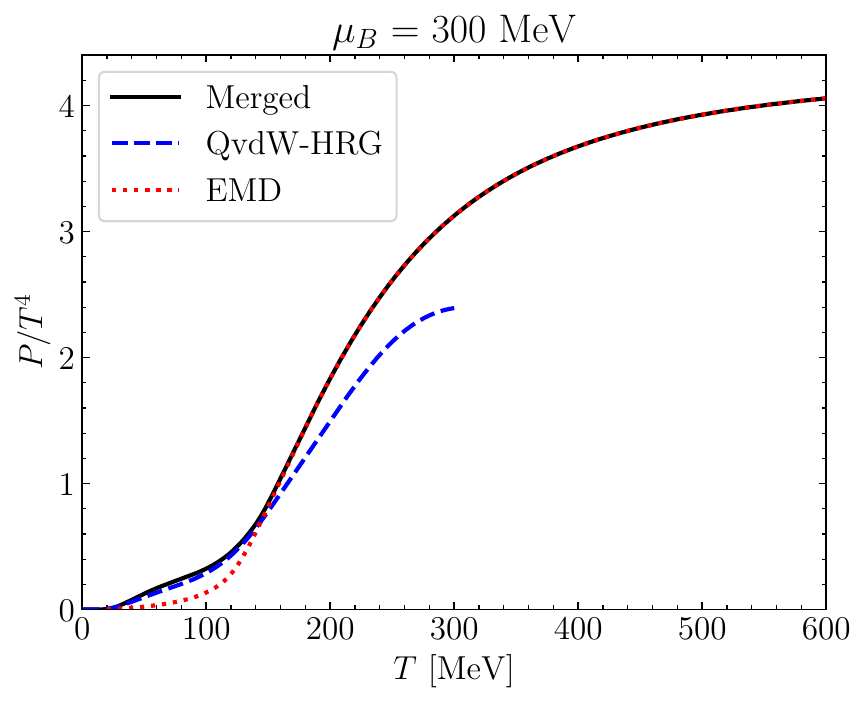}
\end{subfigure}
\begin{subfigure}
     \centering
    \includegraphics[width=0.45\linewidth]{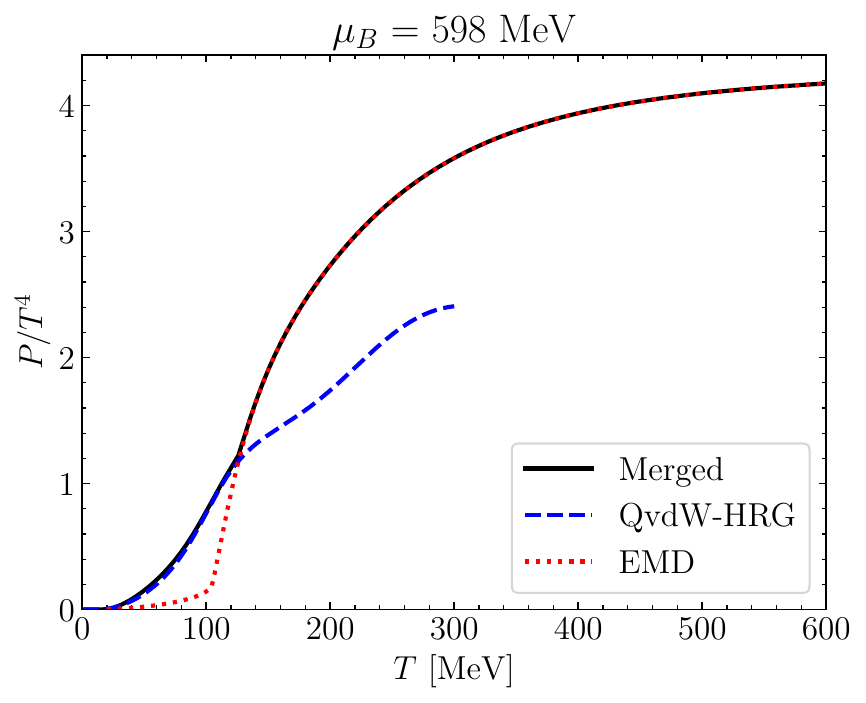}
\end{subfigure}
\begin{subfigure}
     \centering
    \includegraphics[width=0.45\linewidth]{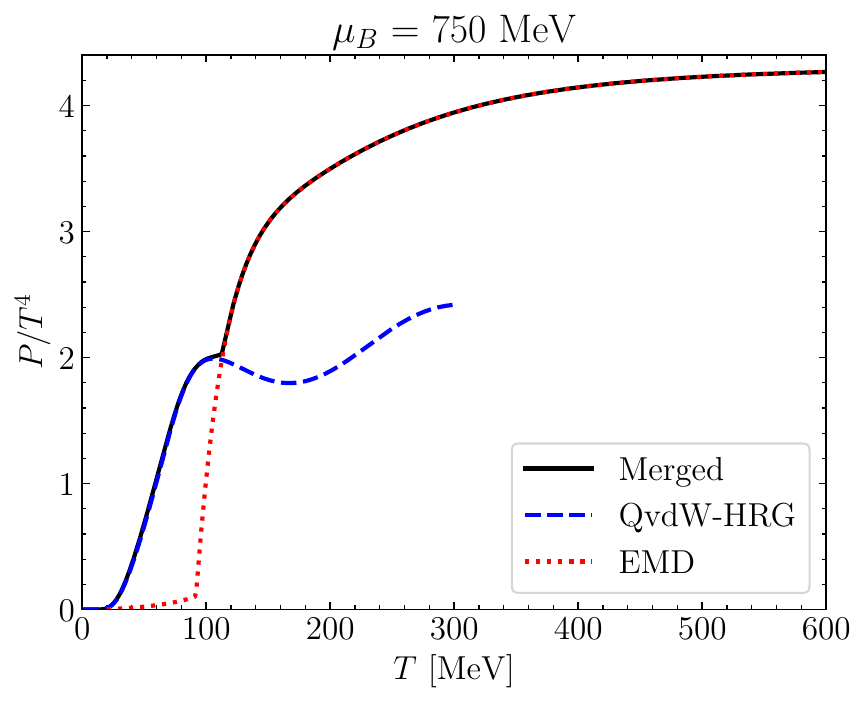}
\end{subfigure}
    \caption{Pressure as a function of the temperature for different values of chemical potential. At chemical potential $\mu_B=0$, we compare the results with LQCD data from Ref. \cite{Borsanyi:2013bia} (green points). In all panels, the solid black line indicates the equation of state obtained by merging the QvdW-HRG EoS at low temperature (blue, dashed line) with the EMD EoS at high temperature (red, dotted line).}
    \label{fig:pressure}
\end{figure}

\begin{figure}[h!]
\begin{subfigure}
     \centering
    \includegraphics[width=0.45\linewidth]{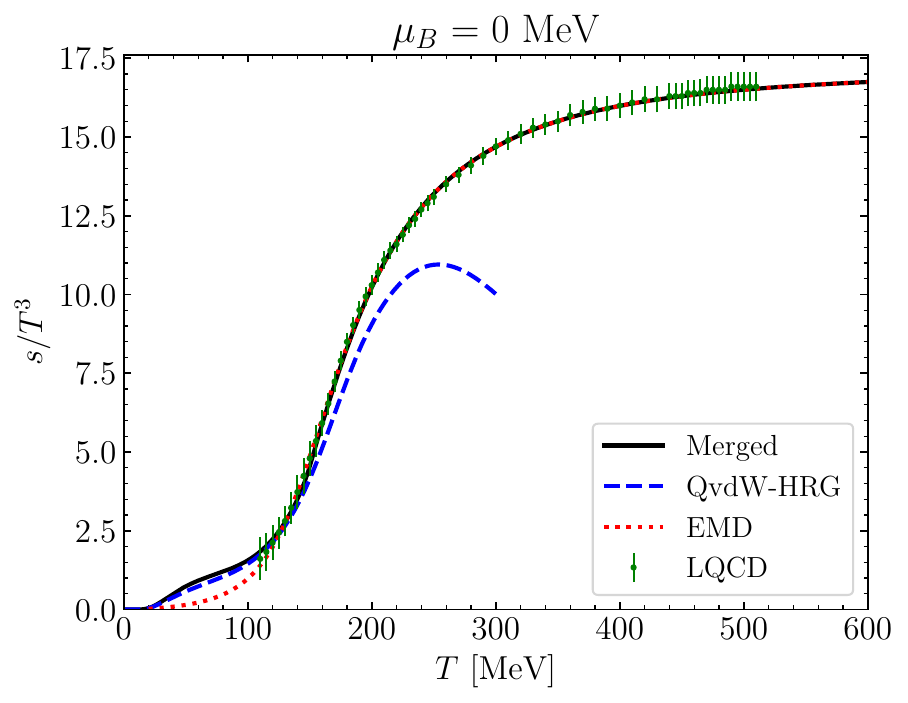}
\end{subfigure}
\begin{subfigure}
     \centering
    \includegraphics[width=0.45\linewidth]{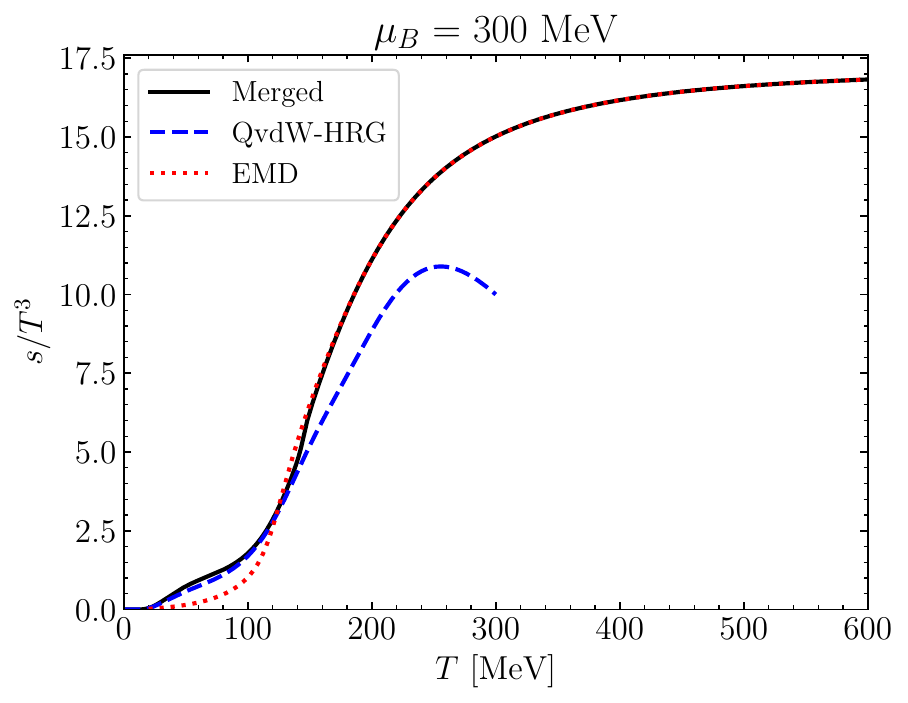}
\end{subfigure}
\begin{subfigure}
     \centering
    \includegraphics[width=0.45\linewidth]{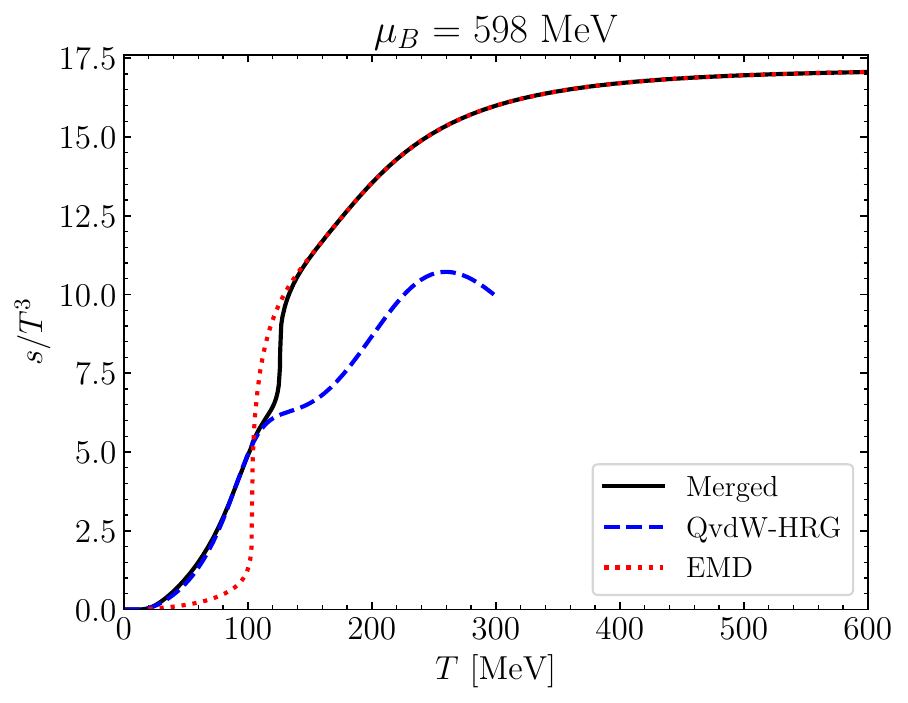}
\end{subfigure}
\begin{subfigure}
     \centering
    \includegraphics[width=0.45\linewidth]{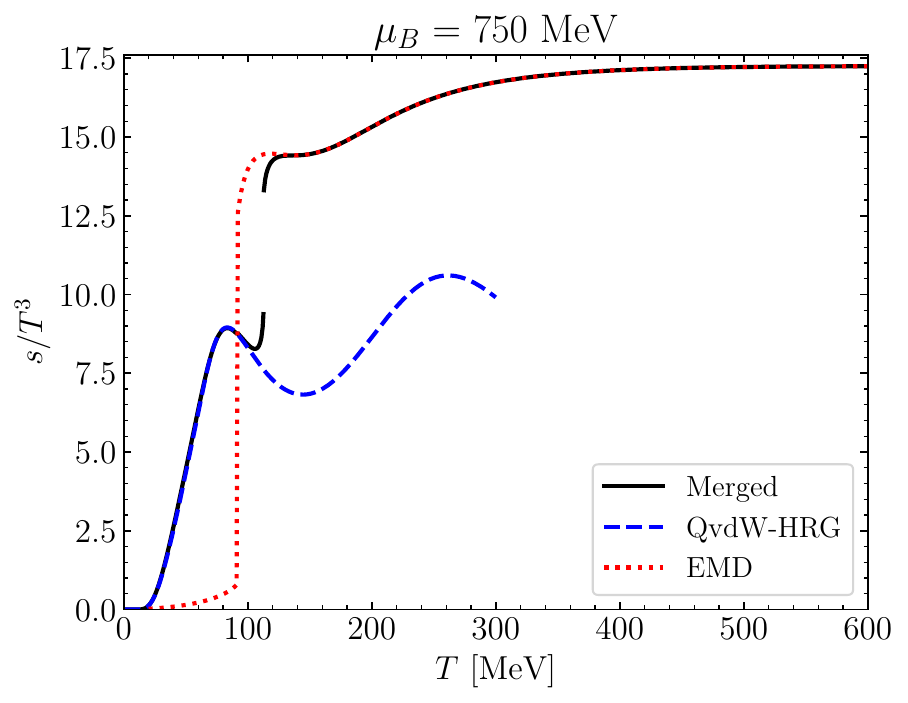}
\end{subfigure}
    \caption{Entropy density as a function of temperature for different values of chemical potential. At vanishing chemical potential $\mu_B=0$, we compare the results with LQCD data from Ref. \cite{Borsanyi:2013bia} (green points). In all panels, the solid black line indicates the equation of state obtained by merging the QvdW-HRG EoS at low temperature (blue, dashed line) with the EMD EoS at high temperature (red, dotted line).}
    \label{fig:entropy}
\end{figure}

\begin{figure}[h!]
\begin{subfigure}
     \centering
    \includegraphics[width=0.45\linewidth]{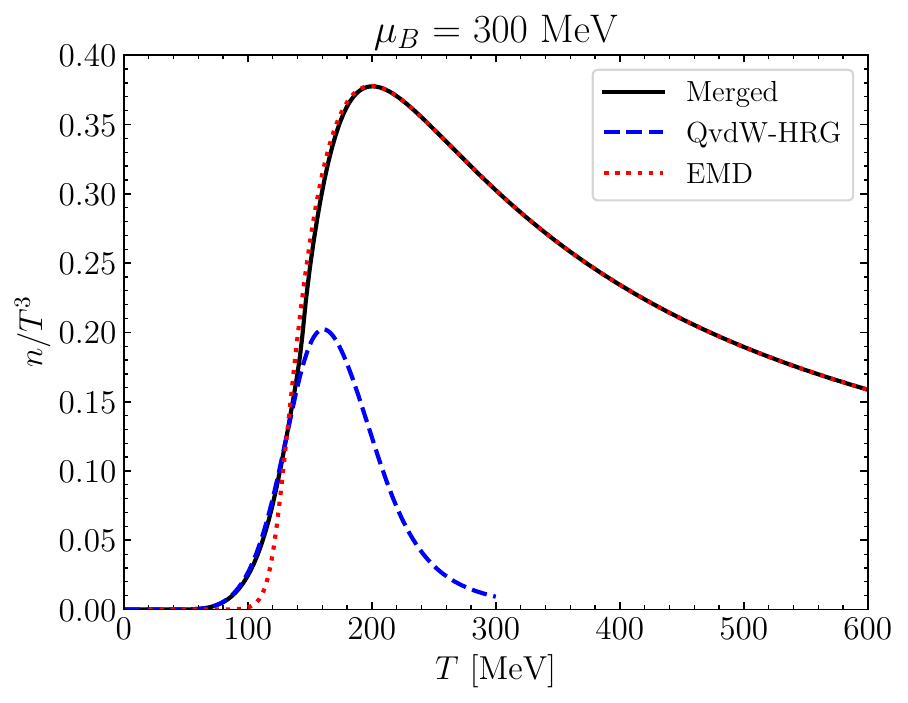}
\end{subfigure}
\begin{subfigure}
     \centering
    \includegraphics[width=0.45\linewidth]{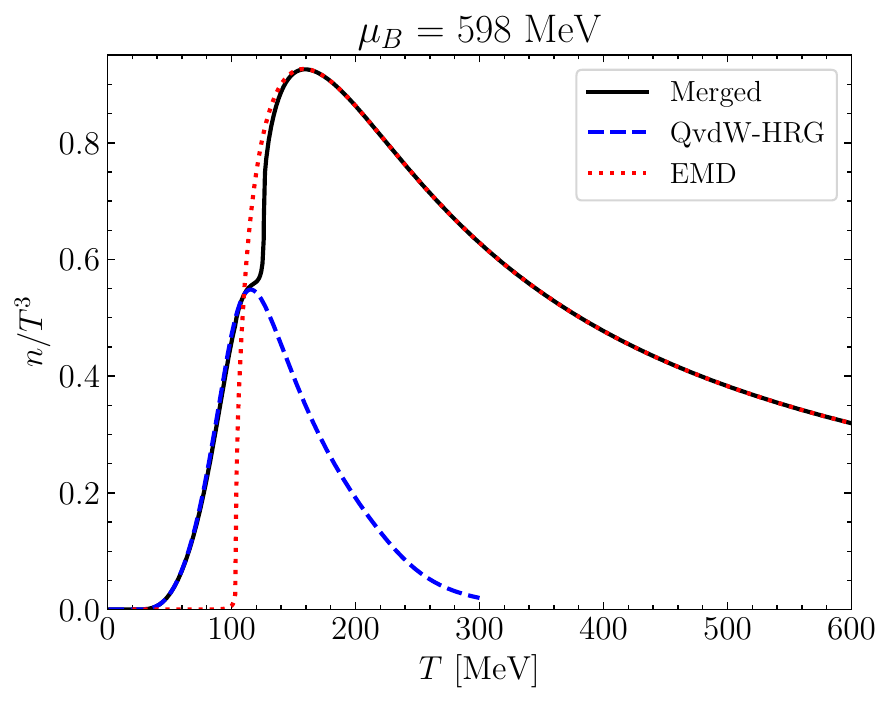}
\end{subfigure}
\begin{subfigure}
     \centering
    \includegraphics[width=0.45\linewidth]{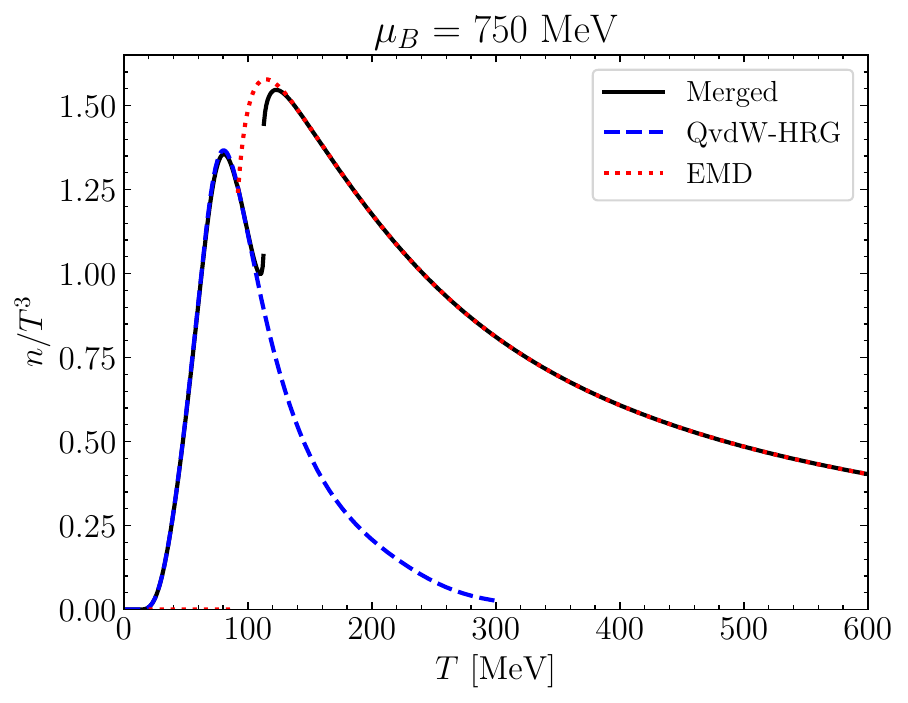}
\end{subfigure}
    \caption{Net baryon density as a function of temperature for different values of chemical potential. In all panels, the solid black line indicates the equation of state obtained by merging the QvdW-HRG EoS at low temperature (blue, dashed line) with the EMD EoS at high temperature (red, dotted line).}
    \label{fig:density}
\end{figure}

\begin{figure}[h!]
\begin{subfigure}
     \centering
    \includegraphics[width=0.45\linewidth]{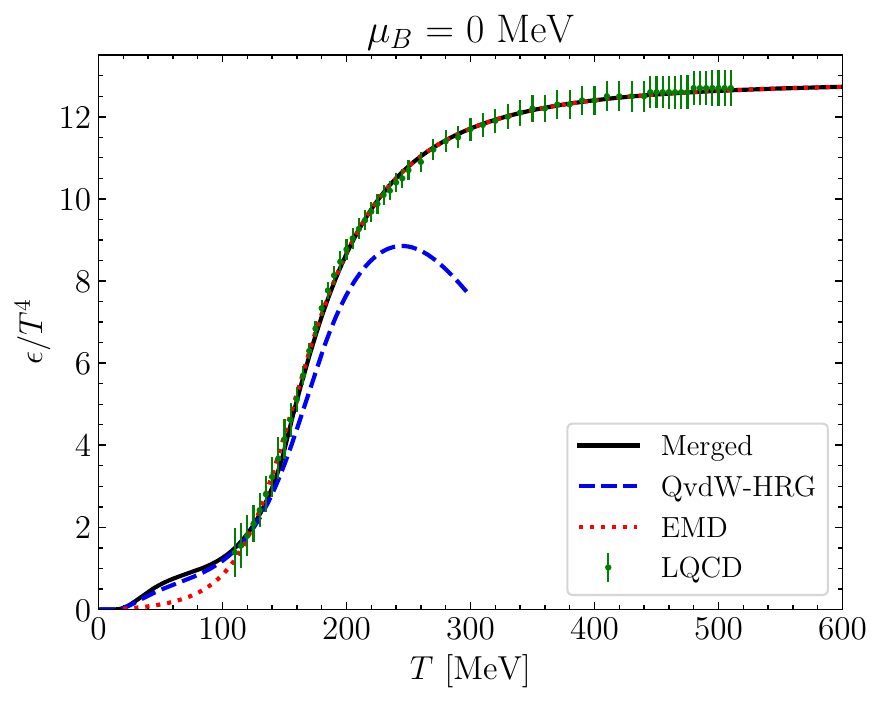}
\end{subfigure}
\begin{subfigure}
     \centering
    \includegraphics[width=0.45\linewidth]{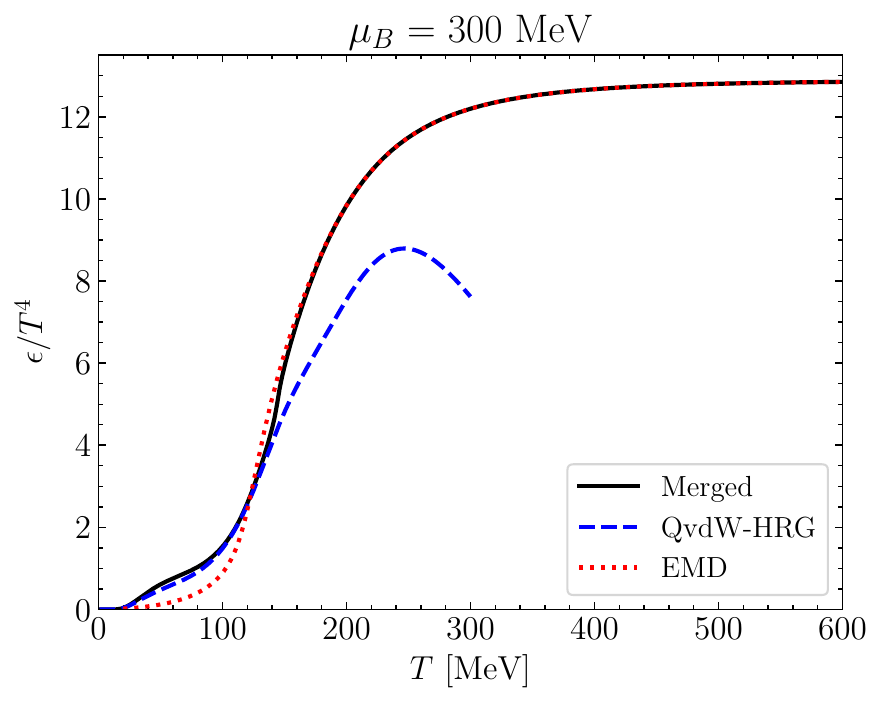}
\end{subfigure}
\begin{subfigure}
     \centering
    \includegraphics[width=0.45\linewidth]{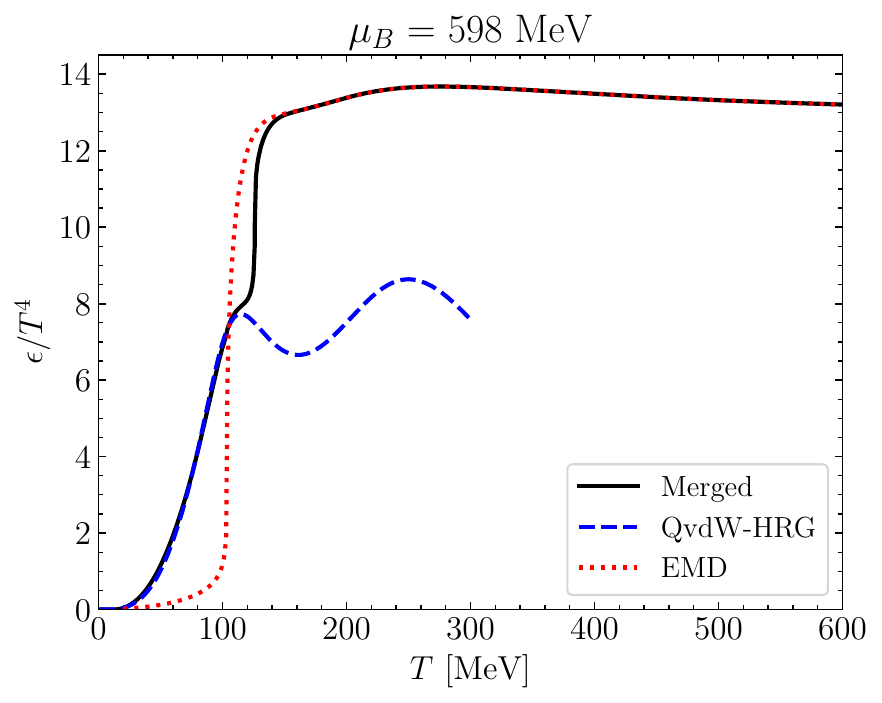}
\end{subfigure}
\begin{subfigure}
     \centering
    \includegraphics[width=0.45\linewidth]{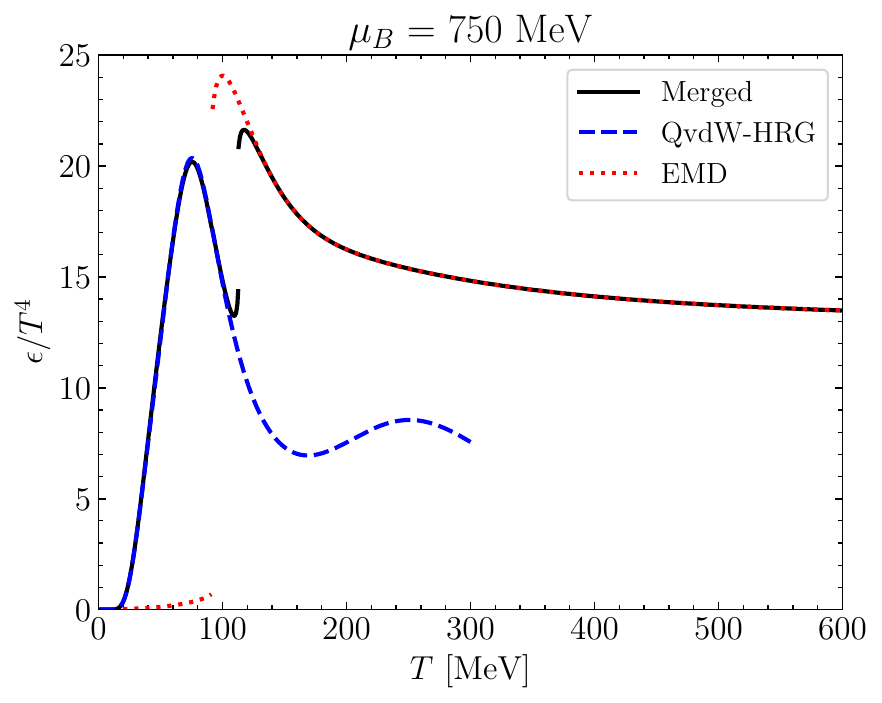}
\end{subfigure}
    \caption{Energy density as a function of temperature for different values of chemical potential.  At vanishing chemical potential $\mu_B=0$, we compare the results with LQCD data from Ref. \cite{Borsanyi:2013bia} (green points). In all panels, the solid black line indicates the equation of state obtained by merging the QvdW-HRG EoS at low temperature (blue, dashed line) with the EMD EoS at high temperature (red, dotted line).}
    \label{fig:endensity}
\end{figure}

\begin{figure}[h!]
\begin{subfigure}
     \centering
    \includegraphics[width=0.45\linewidth]{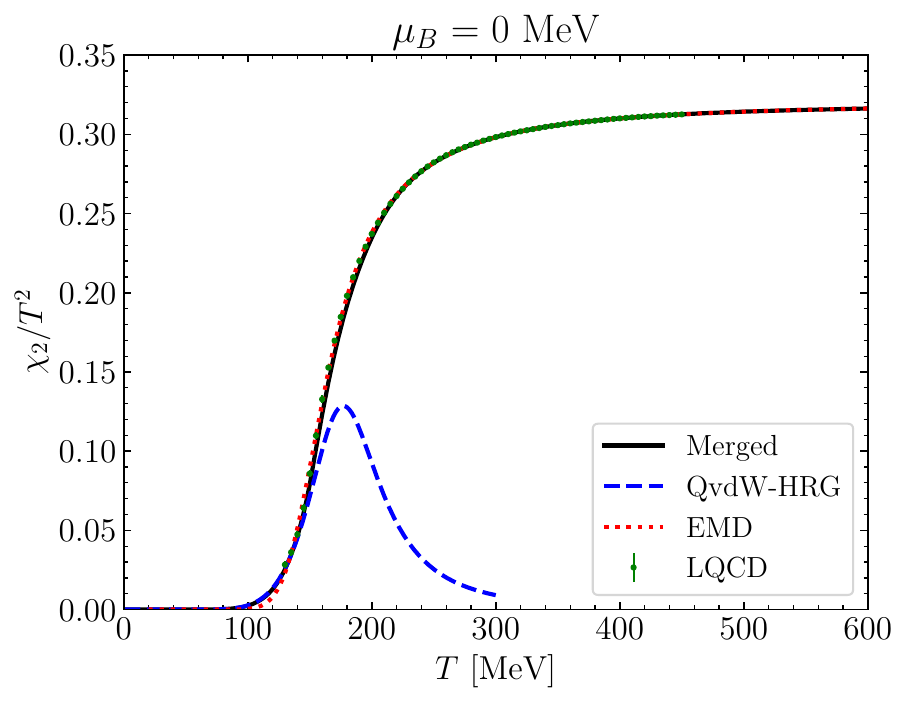}
\end{subfigure}
\begin{subfigure}
     \centering
    \includegraphics[width=0.45\linewidth]{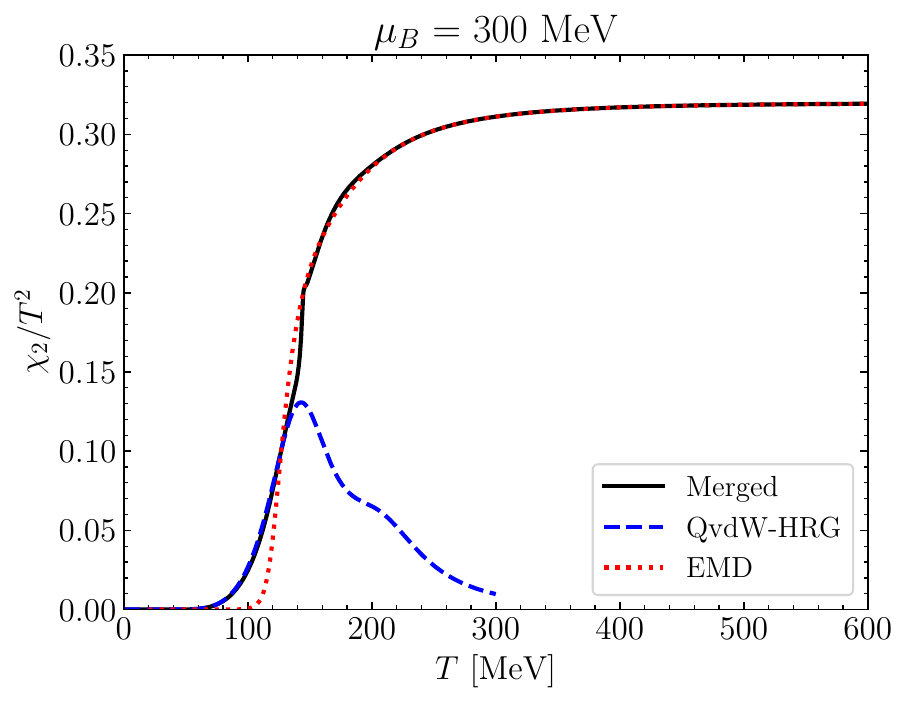}
\end{subfigure}
\begin{subfigure}
     \centering
    \includegraphics[width=0.45\linewidth]{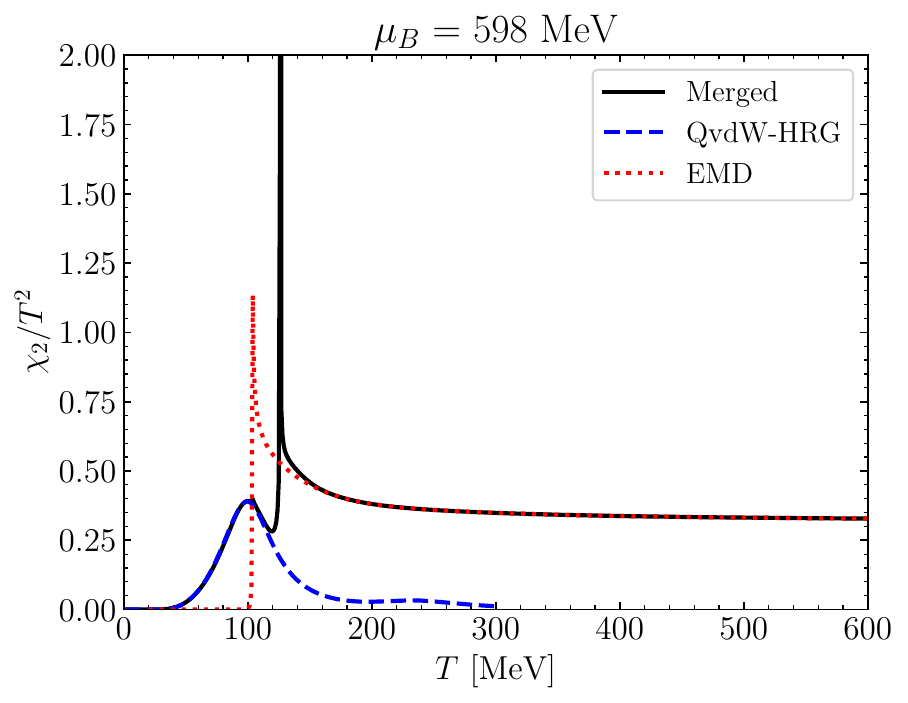}
\end{subfigure}
\begin{subfigure}
     \centering
    \includegraphics[width=0.45\linewidth]{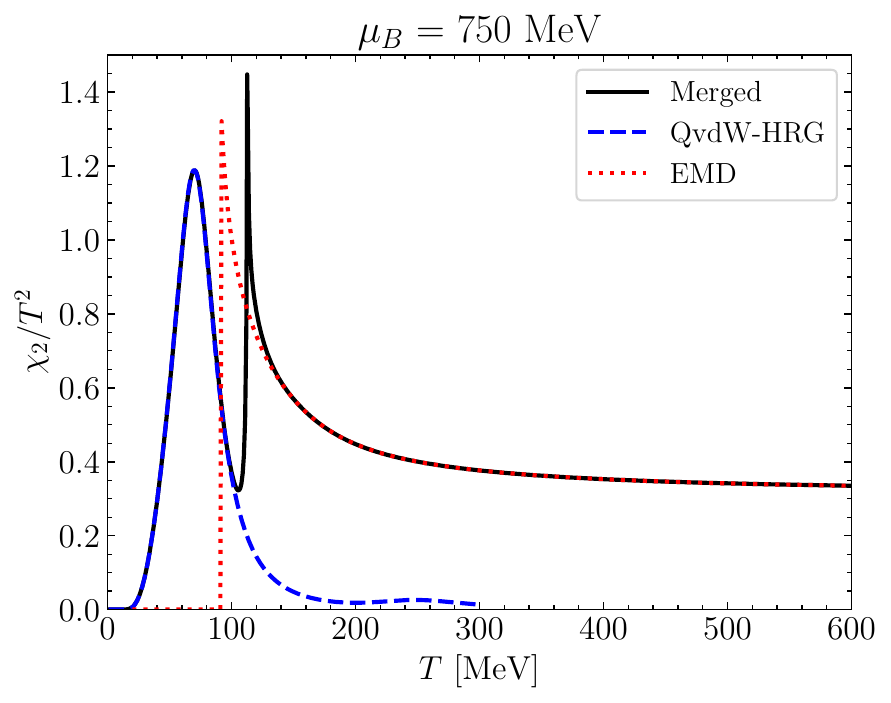}
\end{subfigure}
    \caption{Net baryon number susceptibility as a function of temperature for different values of chemical potential.  At vanishing chemical potential $\mu_B=0$, we compare the results with LQCD data from Ref. \cite{Borsanyi:2021sxv} (green points). In all panels, the solid black line indicates the equation of state obtained by merging the QvdW-HRG EoS at low temperature (blue, dashed line) with the EMD EoS at high temperature (red, dotted line).}
    \label{fig:chi2}
\end{figure}

\begin{figure}[h!]
\begin{subfigure}
     \centering
    \includegraphics[width=0.45\linewidth]{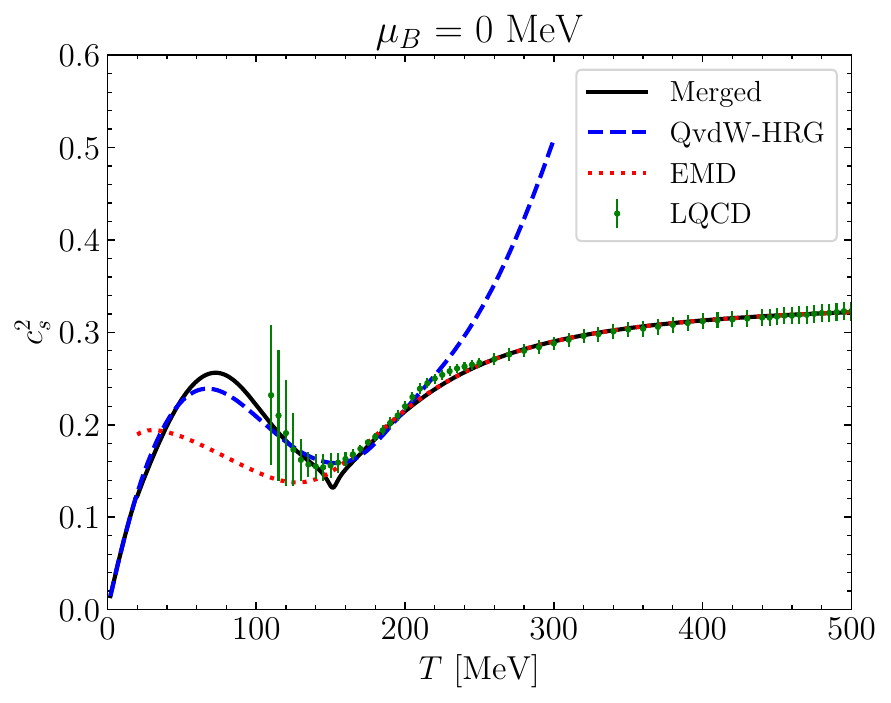}
\end{subfigure}
\begin{subfigure}
     \centering
    \includegraphics[width=0.45\linewidth]{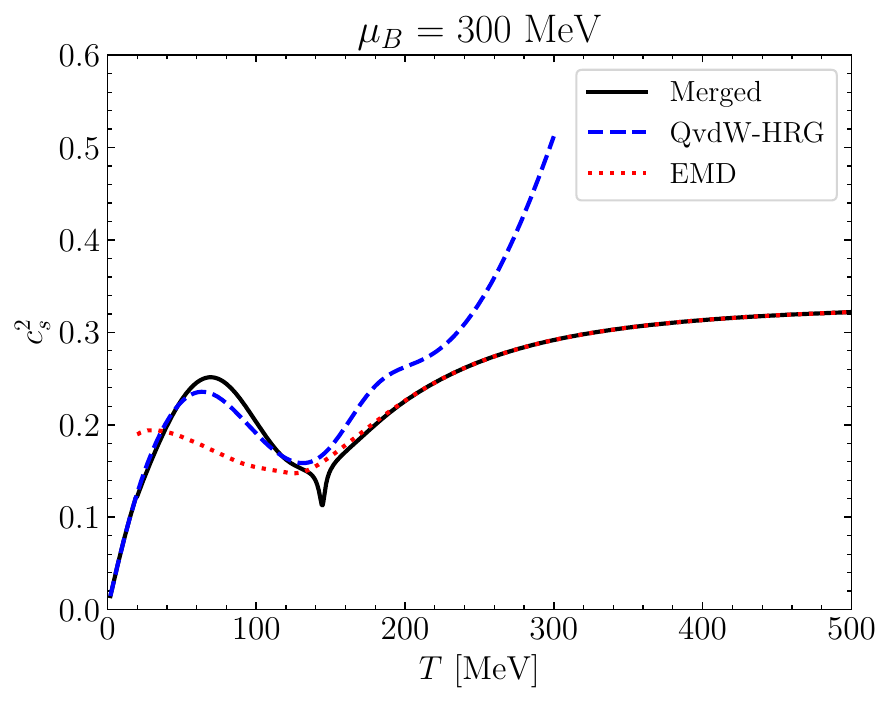}
\end{subfigure}
\begin{subfigure}
     \centering
    \includegraphics[width=0.45\linewidth]{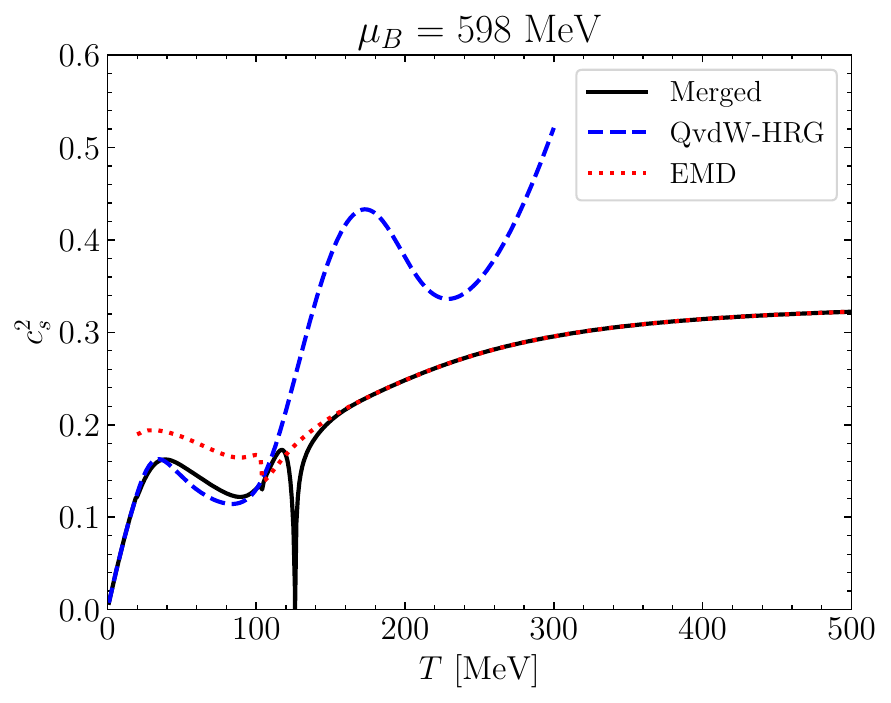}
\end{subfigure}
\begin{subfigure}
     \centering
    \includegraphics[width=0.45\linewidth]{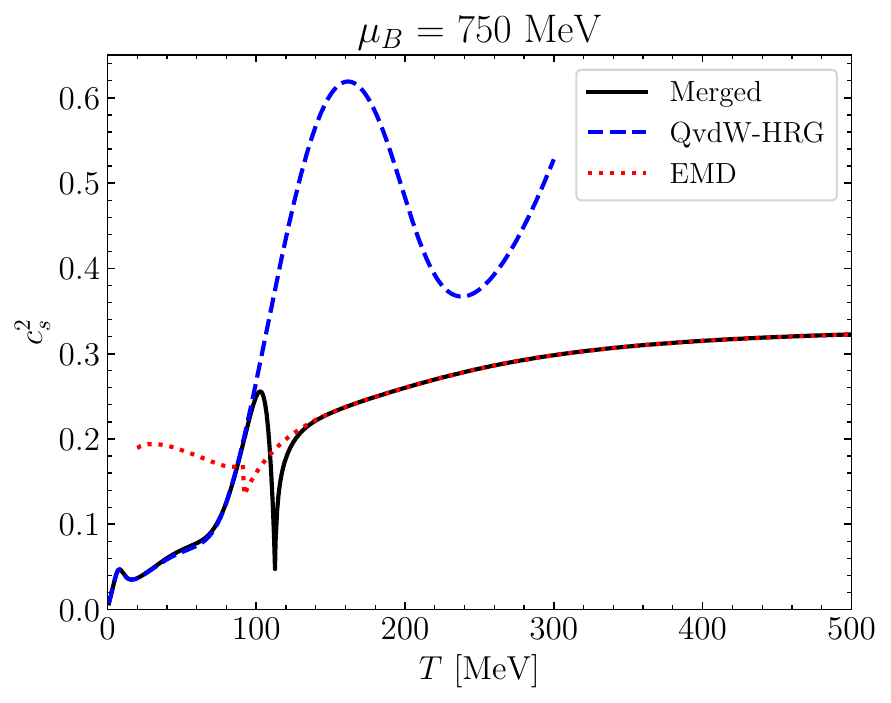}
\end{subfigure}
    \caption{Speed of sound squared as a function of the temperature for different values of chemical potential. At vanishing chemical potential $\mu_B=0$, we compare the results with LQCD data from Ref. \cite{Borsanyi:2013bia} (green points). In all panels, the solid black line indicates the equation of state obtained by merging the QvdW-HRG EoS at low temperature (blue, dashed line) with the EMD EoS at high temperature (red, dotted line).}
    \label{fig:cs}
\end{figure}

Figures~\ref{fig:pressure} to \ref{fig:cs} present thermodynamic quantities at different fixed values of $\mu_B$, shown for the merged EoS as well as for the QvdW-HRG and EMD models. 
To make sure our results are consistent with LQCD, we also compare them to results from \cite{Borsanyi:2013bia,Borsanyi:2021sxv} at  $\mu_B=0$. 

The pressure is shown in Fig. \ref{fig:pressure}, where we verify that the merged pressure closely follows the EoS with the largest value of the pressure for all values of $\mu_B$. 
For $\mu_B>\mu_c$, a kink appears in the merged pressure, as seen in the bottom-left panel, signaling the first-order transition.


Figures~\ref{fig:entropy}, \ref{fig:density}, and \ref{fig:endensity} show the entropy density $s$, net baryon density $n$, and energy density $\epsilon$, respectively, as functions of $T$. 
For $\mu_B<\mu_c$,  the merged EoS continuously transitions from the QvdW-HRG to the EMD EoS as $T$ increases. 
As these are all extensive quantities, we find that they develop a discontinuity for  $\mu_B>\mu_c$, signaling a first-order phase transition. 
We also note that entropy and energy densities agree quite well with LQCD at zero density.


Results for the baryon susceptibility, $\chi_2$, and the speed of sound squared, $c_s^2$, are presented in Figs.~\ref{fig:chi2} and \ref{fig:cs}, respectively.
In that case, we observe slight deviations from LQCD results at $\mu_B=0$ and intermediate values of $T\approx 140 - 180$ MeV. 
This is due to contributions from ${\partial \overline{p}}/{\partial T}$ and ${\partial \overline{p}}/{\partial \mu_B}$ to second derivatives of the pressure, explicitly shown in Eq.~\eqref{eq:2nd_deriv}. 

\begin{figure}[h!]
\begin{subfigure}
     \centering
    \includegraphics[width=0.45\linewidth]{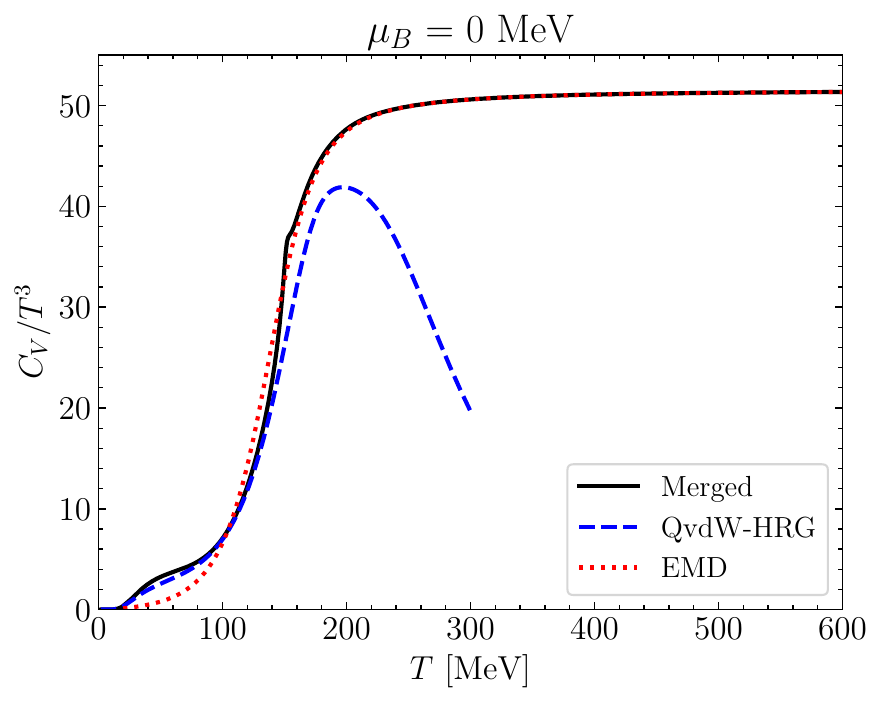}
\end{subfigure}
\begin{subfigure}
     \centering
    \includegraphics[width=0.45\linewidth]{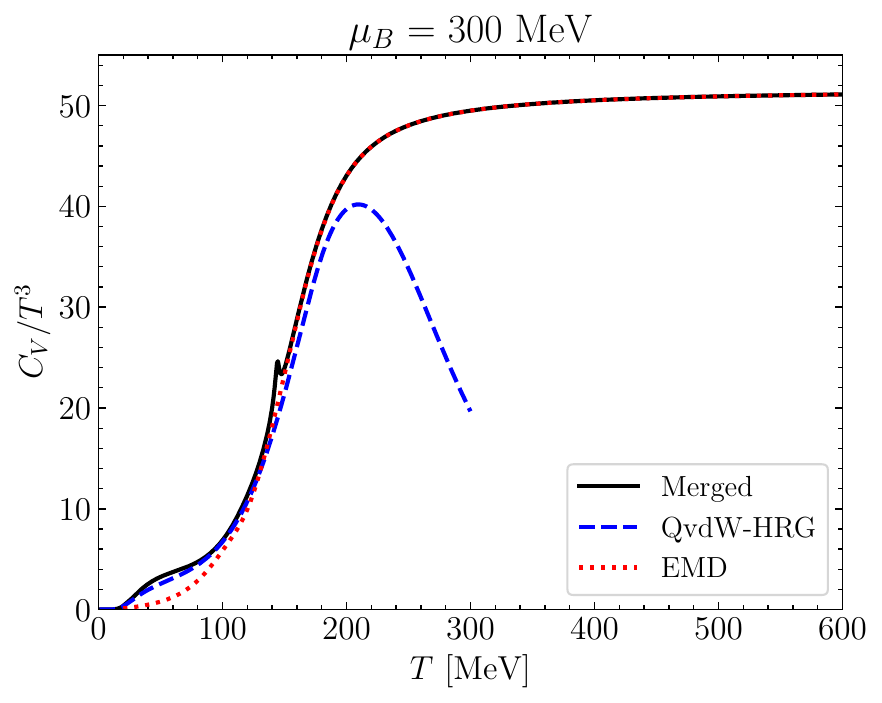}
\end{subfigure}
\begin{subfigure}
     \centering
    \includegraphics[width=0.45\linewidth]{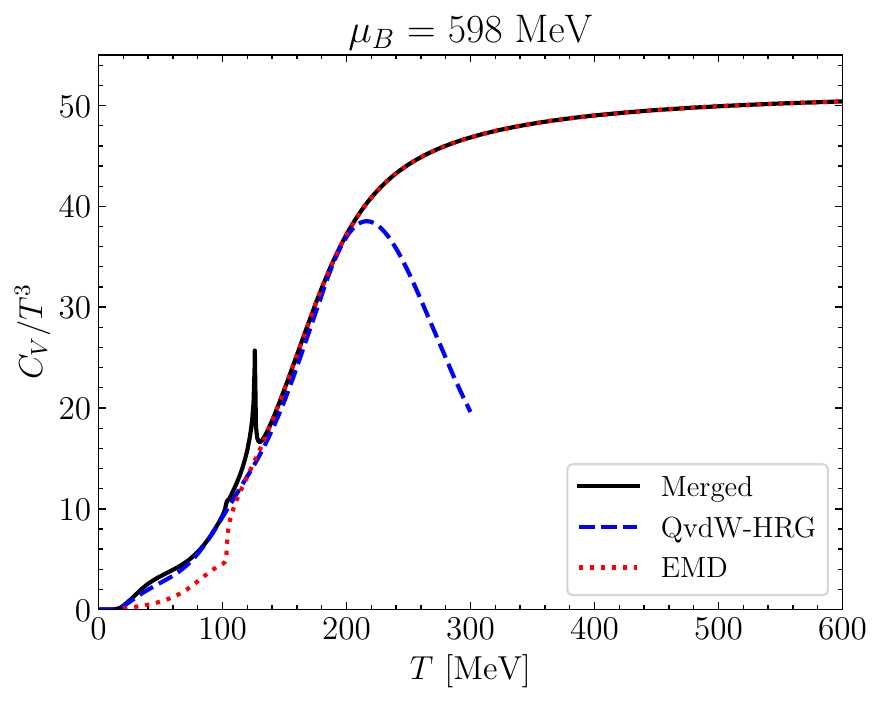}
\end{subfigure}
\begin{subfigure}
     \centering
    \includegraphics[width=0.45\linewidth]{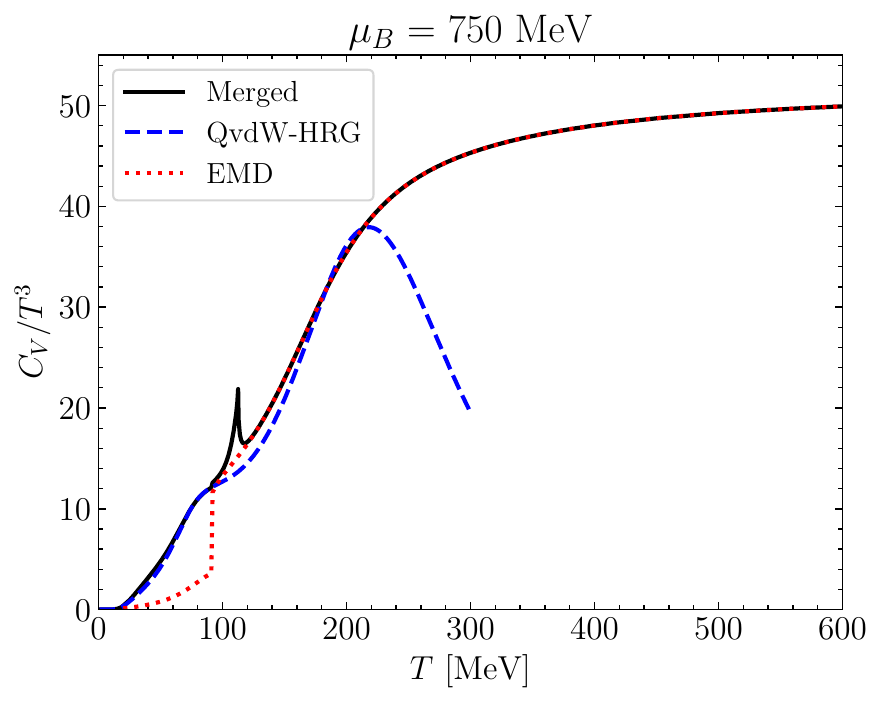}
\end{subfigure}
    \caption{Specific heat at constant volume as a function of temperature for different values of chemical potential. In all panels, the solid black line indicates the equation of state obtained by merging the QvdW-HRG EoS at low temperature (blue, dashed line) with the EMD EoS at high temperature (red, dotted line).}
    \label{fig:cv}
\end{figure}

The effect of these mixing terms becomes more pronounced as $\mu_B$ increases, and at $\mu_B=\mu_c$, they give rise to the divergence in $\chi_2$ and vanishing of $c_s^2$ expected at a critical point. 
Similar behavior is also found for the specific heat at fixed volume, $C_V$, shown in Fig.~\ref{fig:cv}. 
While the mixing contributions to second derivatives are proportional to $\Delta V$, decreasing its value would lead to a larger mixing contribution to the entropy density at low temperatures, which may also cause the merged EoS to deviate from the input models and from LQCD results. 
In general, the value of the phenomenological parameter $\Delta V$ 
must be carefully chosen based on the global analysis of the output in the $T-\mu_B$ plane, ensuring that the merged EoS follows the correct EoS in its respective validity range while keeping merging artifacts in $\chi_2$, $c_s^2$, and $C_V$ under control. 

\begin{figure}[h!]

\centering
     \includegraphics[width=.6\textwidth]{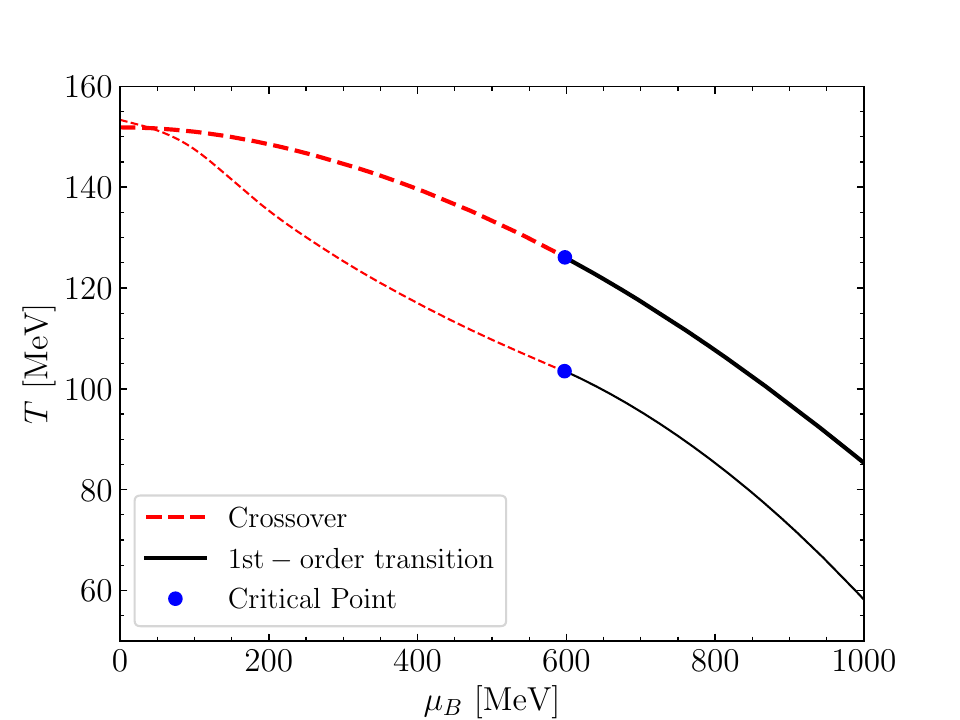}
\caption{Phase diagram in the $T$--$\mu_B$ plane from the merged EoS (thick lines) and from the EMD model (thin lines). The crossover line (red dashed lines) is defined from the condition that $\overline{p}=1/2$, for the merged EoS, and as the inflection point of the baryon susceptibility for the EMD model EoS. 
The solid black lines mark the first-order phase transition, while the dot signals the predicted critical point. 
}\label{fig:phase_diagram}
\end{figure}

 In Fig.~\ref{fig:phase_diagram}, we compare the crossover and first-order lines extracted from the merging procedure  with the ones obtained from the EMD EoS.  
We note that, as there is no sharp distinction between phases without a proper phase transition, the crossover line can vary depending on arbitrary definitions. 
In our merged EoS, it corresponds to the line where the pressures of the two input EoSs intersect, while for the EMD model, it  was found from the inflection point of $\chi_2$ \cite{Grefa:2021qvt}. 
We also stress that the first-order line for our merged EoS is way above the one predicted by the EMD model. 
This is to be expected, since the QvdW-HRG leads to a much higher pressure in the hadronic phase when compared to the holographic model. 
Since the phase transition is a result of a competition between the hadron gas and the QGP, it should be shifted to higher $T$ and $\mu_B$ due to this increase in the hadronic pressure, as is indeed the case.

\begin{figure}[h!]
\begin{subfigure}
     \centering
    \includegraphics[width=0.45\linewidth]{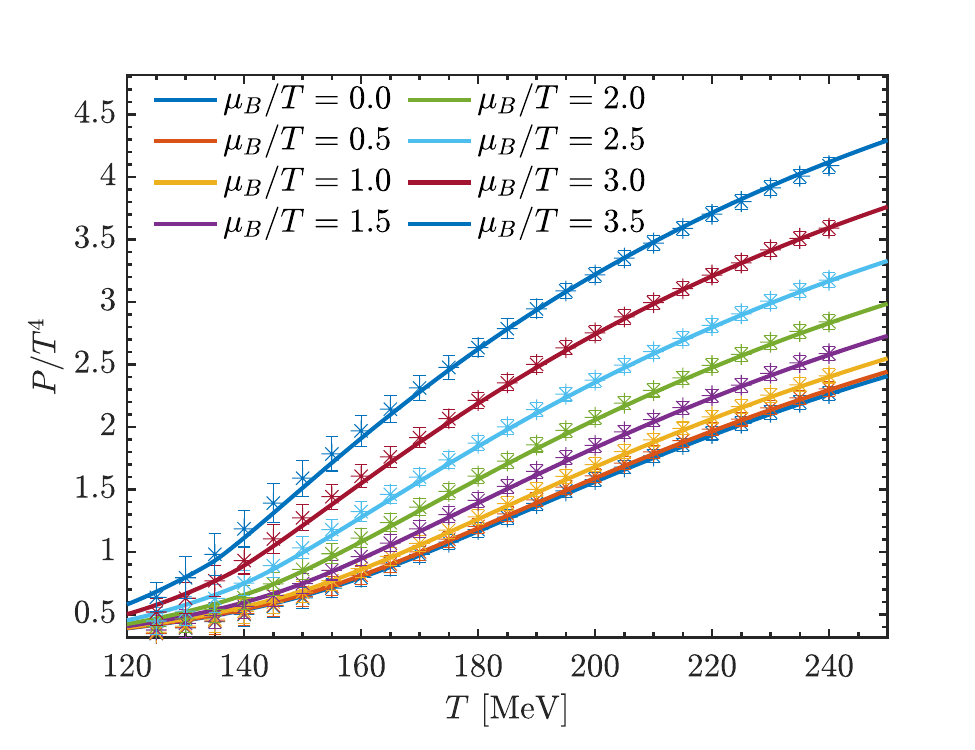}
\end{subfigure}
\begin{subfigure}
     \centering
    \includegraphics[width=0.45\linewidth]{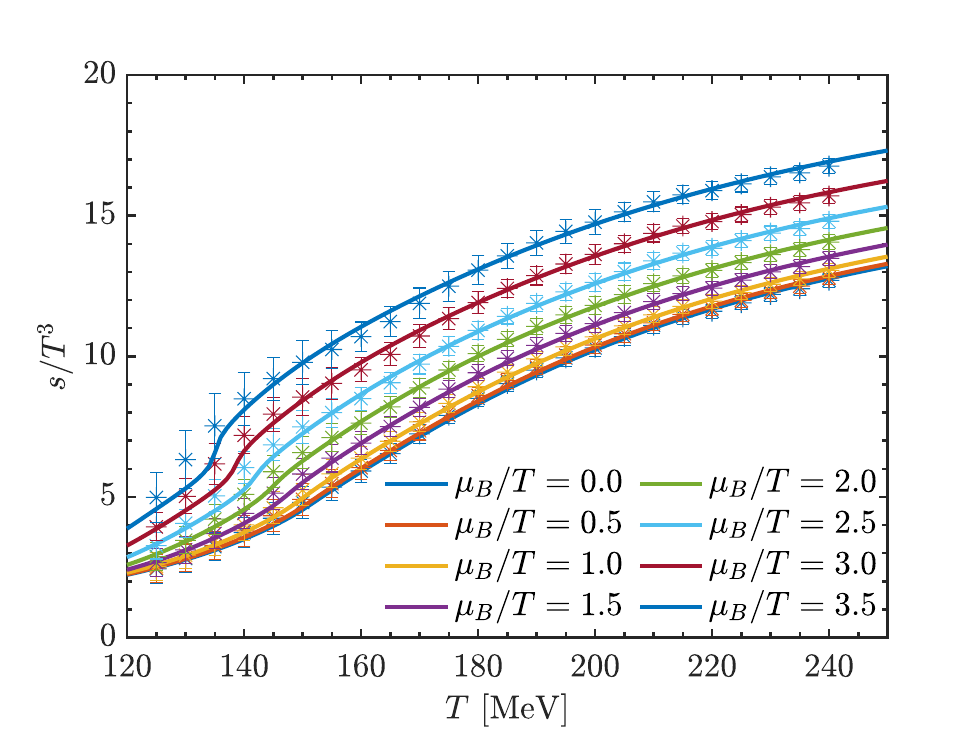}
\end{subfigure}
\begin{subfigure}
     \centering
    \includegraphics[width=0.45\linewidth]{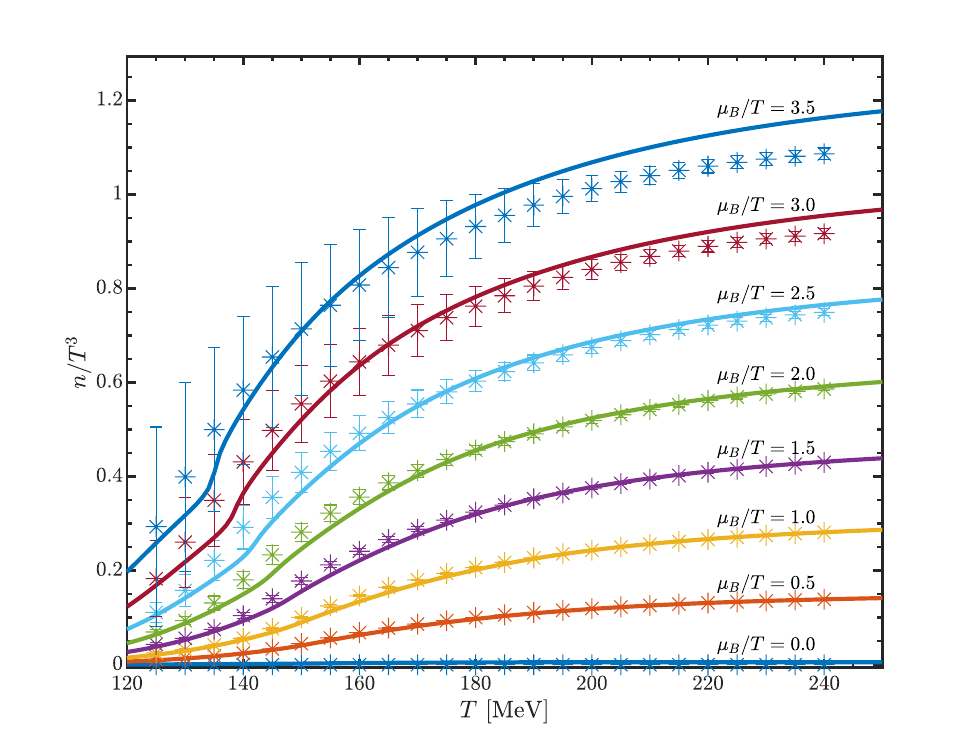}
\end{subfigure}
\begin{subfigure}
     \centering
    \includegraphics[width=0.45\linewidth]{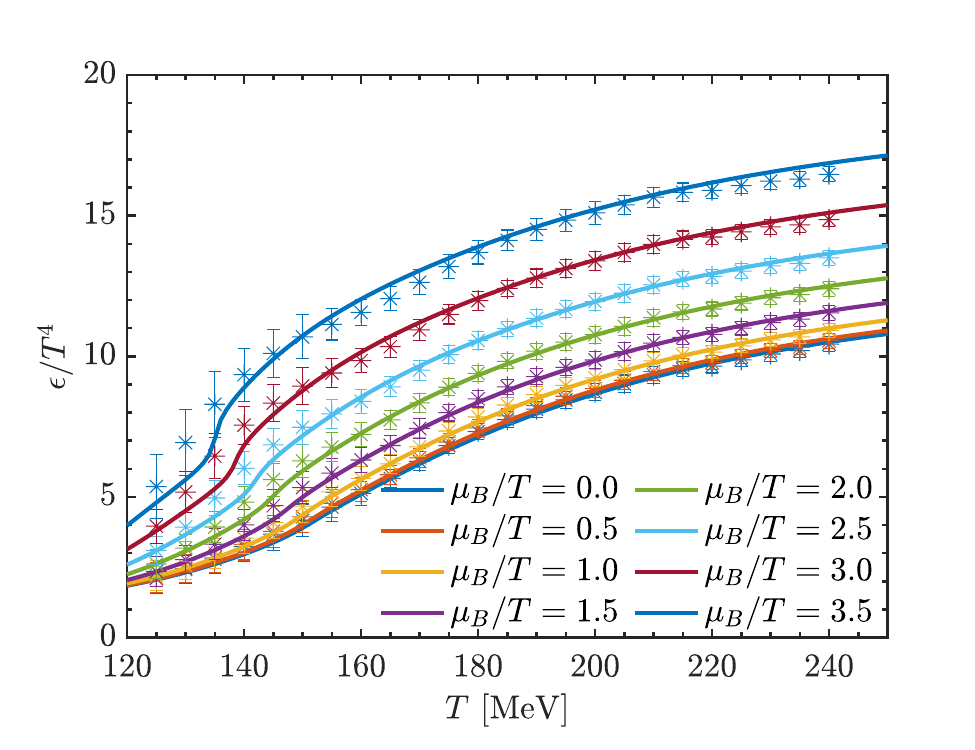}
\end{subfigure}
    \caption{Comparison between the merged EoS (solid curves) and results from the LQCD $T'$-expansion from  Ref.~\cite{Borsanyi:2021sxv} (points with error bars). 
    Results for the pressure (top left), entropy density (top right), baryon density (bottom left), and energy density (bottom right) are shown as functions of $T$ for different values of $\mu_B/T$, shown with different colors.
    }
    \label{fig:lat_comp}
\end{figure}

To validate our results at finite density, we compare them to the EoS obtained from LQCD using the $T'$ expansion from Ref.~\cite{Borsanyi:2021sxv}. 
Figure~\ref{fig:lat_comp} shows comparisons for the pressure, entropy, net baryon density, and energy density, as functions of $T$, at different values of $\mu_B/T$. 
All merged quantities show very good agreement with the LQCD results. 
This could, of course, be anticipated, since both QvdW-HRG and EMD agree quite well with LQCD in their respective range of validity.

\begin{figure}[h!]
\vspace{-0.5in}
\begin{subfigure}
     \centering
\includegraphics[width=0.45\linewidth]{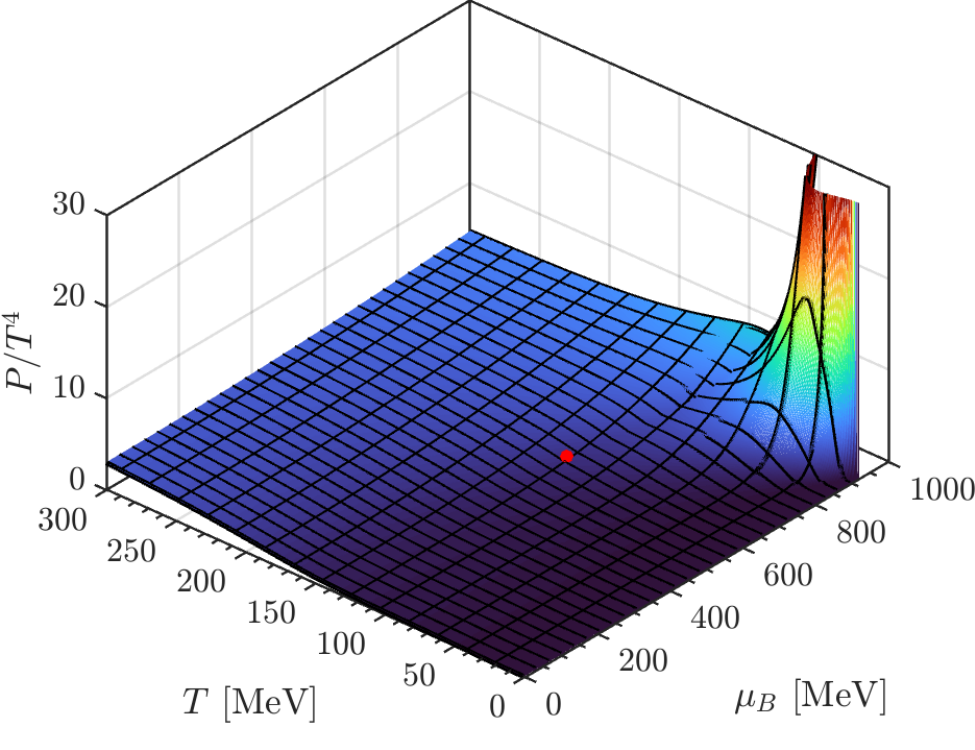}
\vspace{-0.1in}
\end{subfigure}
\begin{subfigure}
     \centering
\includegraphics[width=0.45\linewidth]{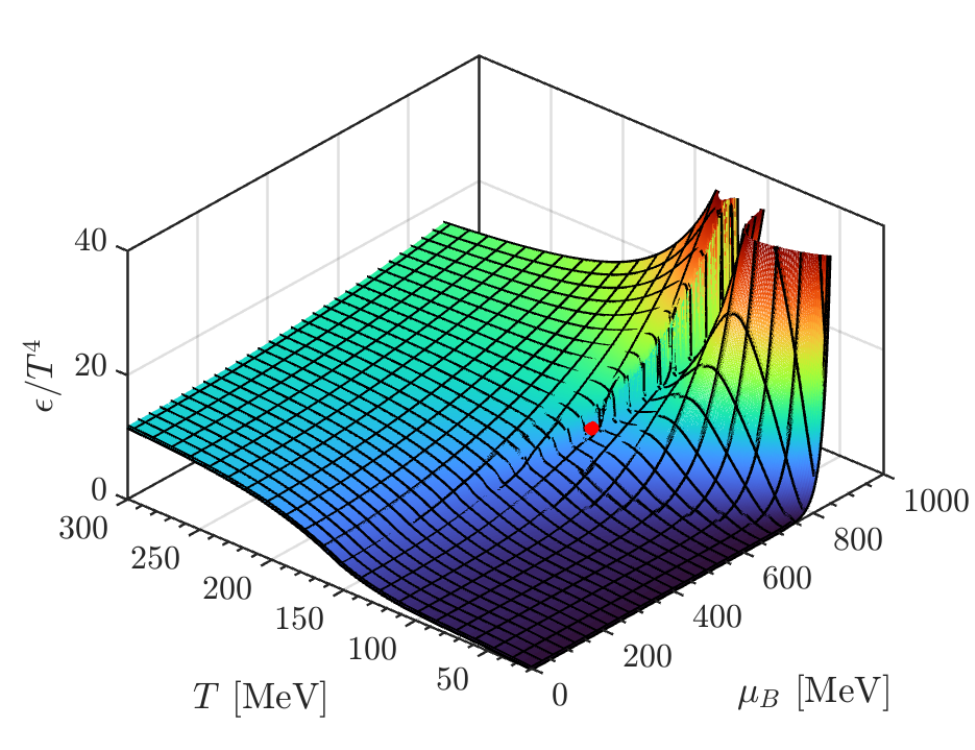}
\end{subfigure}
\begin{subfigure}
     \centering
\includegraphics[width=0.45\linewidth]{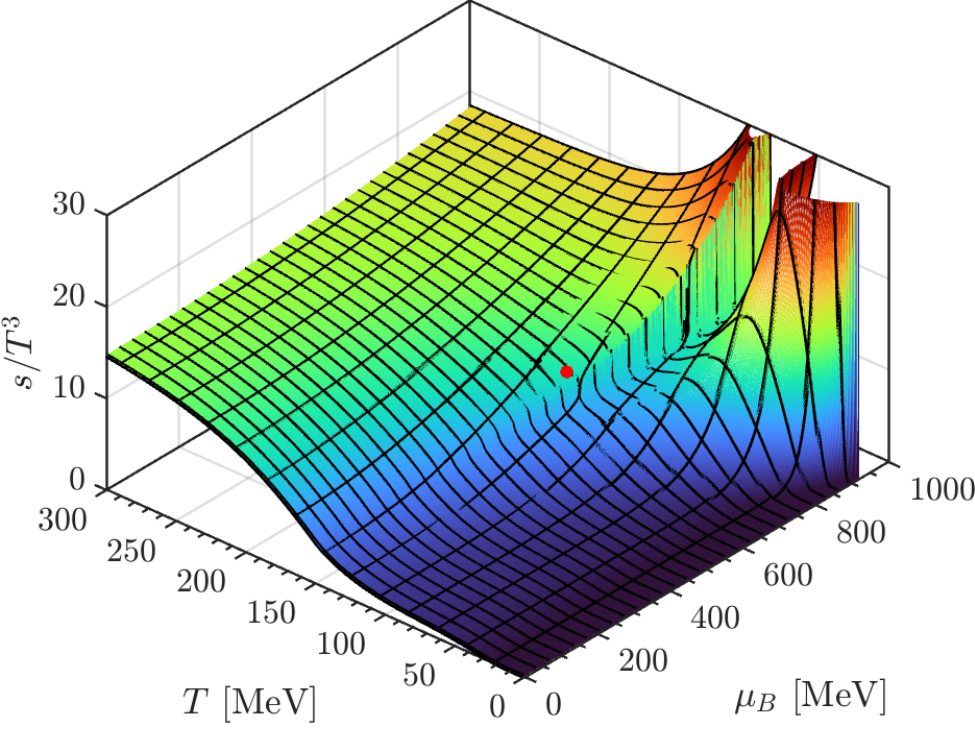}
\end{subfigure}
\begin{subfigure}
     \centering
\includegraphics[width=0.45\linewidth]{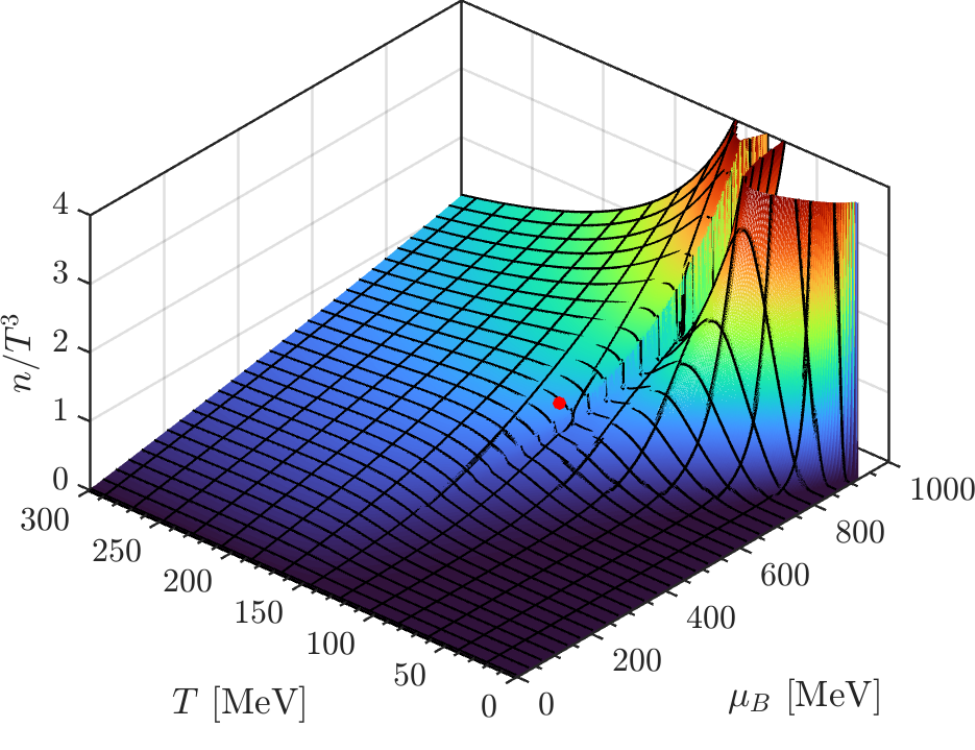}
\end{subfigure}
\begin{subfigure}
     \centering
\includegraphics[width=0.45\linewidth]{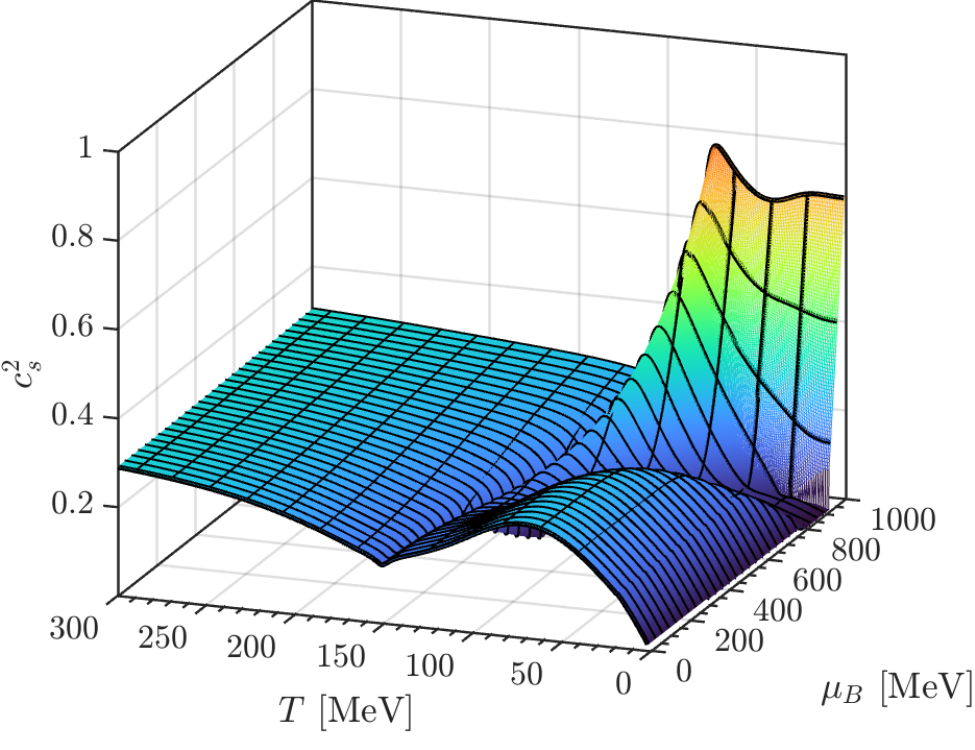}
\end{subfigure}
\begin{subfigure}
\centering
\includegraphics[width=0.45\linewidth]{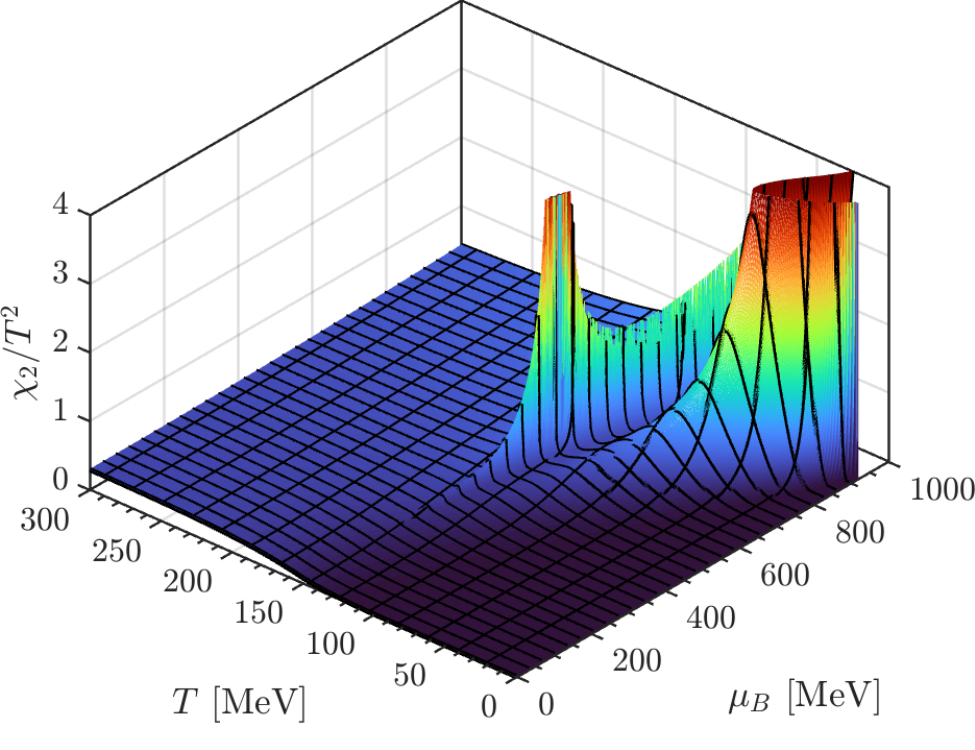}
\end{subfigure}
\begin{subfigure}
\centering
\includegraphics[width=0.45\linewidth]{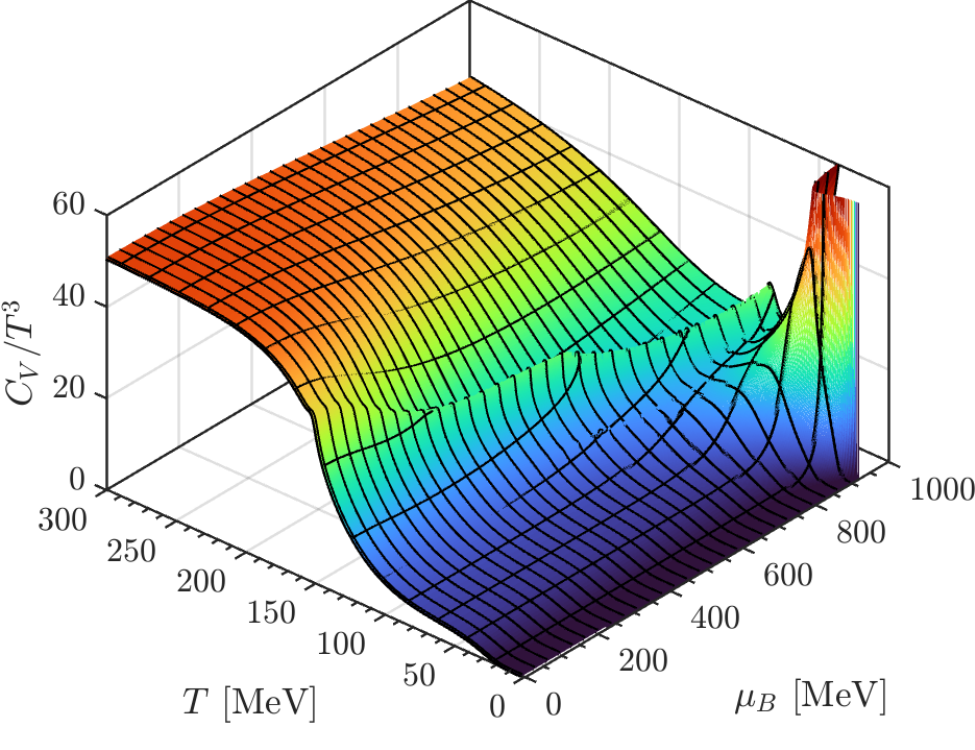}
\end{subfigure}

\vspace{-0.2in}
    \caption{Surface plots of the pressure (top left), energy density (top right), entropy density (center left), net baryon density (center right),  speed of sound squared (bottom left), second order baryon susceptibility  (bottom right), and specific heat at constant volume (last panel on the bottom) as functions of temperature and chemical potential.}
    \label{fig:3Dplots}
\end{figure}
\begin{figure}[h!]
     \centering
    \includegraphics[width=0.7\linewidth]{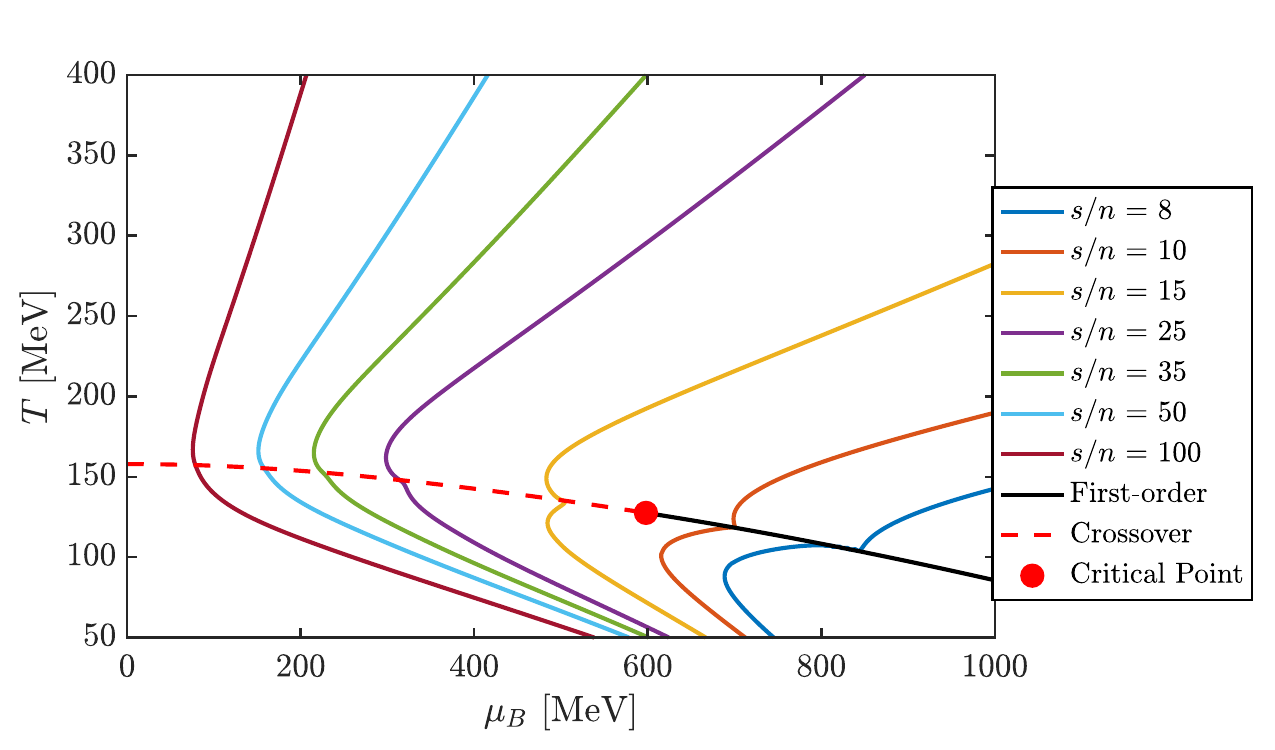}
    \caption{Isentropic trajectories from the merged EoS in the $T$--$\mu_B$ plane. }
    \label{fig:isentropes}
\end{figure}
\begin{figure}[h!]
     \centering
    \includegraphics[width=0.7\linewidth]{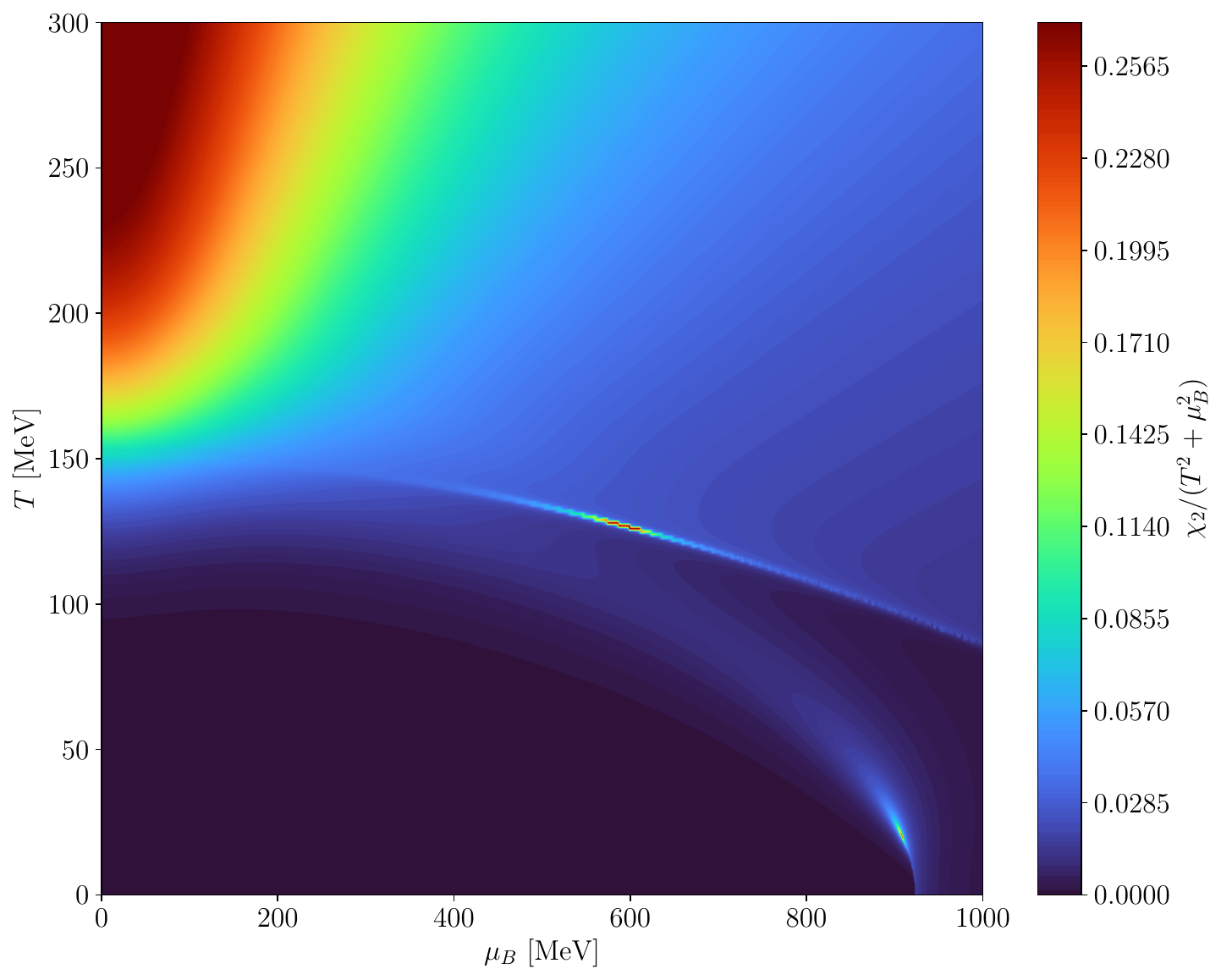}
    \caption{
Heat map of the normalized merged second-order baryon susceptibility 
$\chi_2/(T^2 + \mu_B^2)$, indicating clear signatures of both the nuclear 
liquid--gas critical point at low $T$ and the QCD critical point at 
intermediate $T$ and $\mu_B$.
    \label{fig:chi2contour}}
\end{figure}

In Fig.~\ref{fig:3Dplots}, we provide 3D plots showing thermodynamic quantities as functions of $T$ and $\mu_B$ simultaneously. 
The hallmark features of the critical point and first-order phase transition are clearly visible in this figure, including the discontinuities in extensive quantities, the divergence in the second baryon susceptibility, and the vanishing speed of sound. 

Fig.~\ref{fig:isentropes} shows isentropic trajectories in the $T-\mu_B$ plane. 
At small $\mu_B$, they show the typical behavior expected from the smooth QCD crossover. 
As $\mu_B$ approaches $\mu_c$ from below, these trajectories start bending toward the critical point. 
For chemical potentials larger than the critical one, the first-order line leads to discontinuous trajectories.

Finally, Figure \ref{fig:chi2contour} shows a heat map of the normalized second order baryon susceptibility in the temperature and baryon chemical potential plane. Clearly visible are the change of behavior around the transition line, which goes from smooth to sharp when increasing $\mu_B$, and the narrow peaks around the two critical points for the deconfinement and liquid-gas phase transitions. This quantity diverges at these critical points, as expected.

\section{Conclusions \label{Sec:Conc}}

In this work, we presented a thermodynamically consistent and stable framework to merge two independent model EoSs into a single, global EoS covering a broader region of the QCD phase diagram. 
In contrast to widely used linear switching strategies, which can imprint spurious structures into conserved charge derivatives and even jeopardize convexity, our construction reduces such artifacts by design. 
Our strategy is to promote the weights used in usual linear switching from a prescribed function $S(T,\mu_B)$ to an internal order-parameter-like variable $p\in[0,1]$. 
Hence, at fixed $(T,\mu_B)$, this mixing weight variable is chosen to minimize the 
grand-potential density $\omega(T,\mu_B;p)$, so that $P(T,\mu_B)=-\min_p\omega(T,\mu_B;p)$.

This internal-variable formulation based on a free energy enforces thermodynamic consistency and yields a convex pressure wherever the input EoSs and the free-energy Ansatz are stable. 
Our method also brings an immediate practical advantage concerning numerical robustness: closed-form expressions for the first and second derivatives of $P$, including $s$, $n$, $\chi_2$, and derived quantities such as $c_s^2$ and $C_V$, remove the need for noisy finite-difference procedures and stabilize downstream analysis workflows.

We provide a physically-motivated Ansatz for the free-energy density,  defined by Eq.~\eqref{eq:free_energy}, where the stability of the equilibrium state is ensured as long as $a>0$. 
An interaction term for the mixing weight, $b\,p(1-p)$, enables non-analytic features corresponding to a phase transition while providing us with transparent control over the phase diagram: $b<2a$ yields a crossover, $b=2a$ a critical point, and $b>2a$ a first-order line. 
A $\mathbb{Z}_2$ symmetry for the order parameter $p$, treated at mean-field level, leads to critical exponents in the universality class of the mean-field Ising model. 

As an application of our new method, we merged a QvdW-HRG EoS, describing the hadronic phase, with a nonconformal holographic EMD EoS, modeling a strongly coupled QGP. 
We obtained a thermodynamically consistent and stable EoS that can be used in QCD phase diagram applications spanning up to $\mu_B\sim 1\,\mathrm{GeV}$ and $T\sim 600\,\mathrm{MeV}$, while showing good agreement with LQCD results where they are available, including results for $\mu_B=0$ and from the $T'$ expansion developed in \cite{Borsanyi:2021sxv}. 
The full merged EoS up to $\mu_B= 1000\,\mathrm{MeV}$ and $T= 600\,\mathrm{MeV}$ is publicly available for download at \cite{yang_2026_18176810}. 
This EoS reproduces 
the distinguishing features of a critical point and first-order line---namely, discontinuities of the extensive variables, the divergence of susceptibilities, and a vanishing speed of sound at the critical point. 
Despite the major gains in robustness introduced by our method, we find it still creates small artifacts in $c_s^2$ and $C_V$ at low densities, for $T\approx 140 - 180$ MeV. 
It is possible that such artifacts may be reduced upon changes to our chosen parameters. However, such changes may introduce artifacts in other quantities, so further work is needed in this direction.

The present work establishes a stable, pipeline-ready baseline to merge EoSs in a thermodynamically consistent and stable manner (which includes a critical point). It would be very interesting to use this new procedure to merge other EoSs for the hadronic phase and the deconfined phase  (constrained by LQCD results) to assess the robustness of the features of the merged QCD EoS found in this work. This would be important in hydrodynamic simulations and phenomenological modeling of the baryon-rich QGP formed in low energy heavy-ion collisions at RHIC and in the future FAIR facility. In this regard, we remark that the open-source code for the merging procedure presented in this work is available at \cite{pelicer_2025_17584997} and within the MUSES framework. The Python code specifically designed for this project is available at \cite{yang_2026_18176810}.

Several other extensions of the work presented here are natural. 
First, a systematic uncertainty quantification for $a(T)$ and $T_c$, such as a Bayesian analysis, calibrated simultaneously to LQCD at $\mu_B=0$ and to hadronic constraints, would enable statistically controlled phase-diagram inferences. 
Second, extending the formalism to include electric-charge and strangeness chemical potentials, and to enforce global constraints relevant for heavy-ion collisions and neutron-star matter, would broaden applicability. 
Finally, embedding this EoS into dynamical simulations will allow us to quantify the phenomenological impact of the crossover vs. first-order scenarios on observables sensitive to $\chi_2$ and $c_s^2$. We leave such interesting studies to future work.

\section*{Acknowledgments}{
We thank V.~Vovchenko and H.~Shah for their help obtaining an equation of state for the hadron-resonance gas with the features  required for this work.  
This material is based upon work supported by the National Science Foundation under grants No. PHY-2208724, PHY-2116686 and PHY-2514763, and within the framework of the MUSES collaboration, under grant number No. OAC-2103680. This material is also based upon work supported by the U.S. Department of Energy, Office of Science, Office of Nuclear Physics, under Award Number DE-SC0022023 and by the National Aeronautics and Space Agency (NASA)  under Award Number 80NSSC24K0767.
M.H. was supported by the Brazilian National Council for Scientific and Technological Development (CNPq) under process No. 313638/2025-0. 
Y.Y. and J.N. are partly supported by
the U.S. Department of Energy, Office of Science,
Office for Nuclear Physics under Award No. DE-
SC0023861. R.R. acknowledges financial support by CNPq under grants number 407162/2023-2 and 305466/2024-0.}

\bibliography{bibliography.bib}

\end{document}